\documentclass[nonacm]{acmart}

\AtBeginDocument{%
  }

\usepackage{xspace}
\newcommand{\name}{{RFconstruct}\xspace}

\usepackage{tikz}
\usepackage{afterpage}
\usepackage{multirow}
\usepackage{enumitem}

\setcopyright{none}
\begin{document}

\title{3D Object Reconstruction with mmWave Radars}

\author{Samah Hussein}
\affiliation{%
 \institution{EPFL}
 \city{Lausanne}
 \country{Switzerland}}

 \author{Junfeng Guan*}
 \affiliation{%
  \institution{Bosch Research}
  \country{USA}
  \thanks{*Work done whilst at EPFL}}

\author{Swathi Narashiman}
\affiliation{%
 \institution{EPFL}
 \city{Lausanne}
 \country{Switzerland}}

\author{Saurabh Gupta}
\affiliation{%
 \institution{UIUC}
 \city{Illinois}
 \country{USA}}

\author{Haitham Hassanieh}
\affiliation{%
 \institution{EPFL}
 \city{Lausanne}
 \country{Switzerland}}

\renewcommand{\shortauthors}{Hussein et al.} 
\authorsaddresses{}

\begin{abstract}
This paper presents \name, a framework that enables 3D shape reconstruction using commercial off-the-shelf (COTS) mmWave radars for self-driving scenarios. \name\ overcomes radar limitations of low angular resolution, specularity, and sparsity in radar point clouds through a holistic system design that addresses hardware, data processing, and machine learning challenges. The first step is fusing data captured by two radar devices that image orthogonal planes, then performing odometry-aware temporal fusion to generate denser 3D point clouds. \name\ then reconstructs 3D shapes of objects using a customized encoder-decoder model that does not require prior knowledge of the object's bound box. The shape reconstruction performance of \name\ is compared against 3D models extracted from a depth camera equipped with a LiDAR. We show that \name\ can accurately generate 3D shapes of cars, bikes, and pedestrians.  
\end{abstract}

\settopmatter{printacmref=false}

\maketitle

\section{Introduction}
\label{sec:intro}

Self-driving cars require precise and high-resolution 3D perception of the environment they navigate. The ability to recover accurate depth information, 3D dimensions, and the shapes of objects in the scene can be essential for improving decision-making for safer and more efficient autonomous driving. Today, self-driving cars rely mainly on cameras or LiDAR to image the environment. However, both sensors fail in adverse weather conditions such as fog, smog, snowstorms, and sandstorms~\cite{zang2019impact,norouzian2019snowfall,norouzian2020rain}
which is a foundational challenge against realizing the true vision of autonomous driving. This poses the question: Can we deploy an auxiliary sensor modality that provides high-fidelity 3D information in driving scenarios where optical sensors fail?

Millimeter-wave (mmWave) radars have recently received significant interest in the autonomous driving industry due to their unique ability to operate in adverse weather conditions ~\cite{waymo2021fog,major2019Qualcomm,Rebut_2022_CVPR,meyer2021graph}. Millimeter wave signals can penetrate through fog, smoke, sand, rain, etc., making them resilient to bad weather. They also provide highly accurate depth information and are relatively low cost compared to LiDARs. As a result, many radar manufacturers have extended the capabilities of their autonomous driving radars from 2D ranging to generating 3D point clouds~\cite{smartradar, Smartmicro, arbe, Vayyar, zendar}. Despite this significant progress, there is still a massive gap in the resolution compared to LiDAR and cameras, which prevents mmWave radars from providing detailed and interpretable shapes of objects.

\begin{figure*}
    \centering
    \begin{tikzpicture}
        \node at (0, 0) 
        {\includegraphics[width=\textwidth]{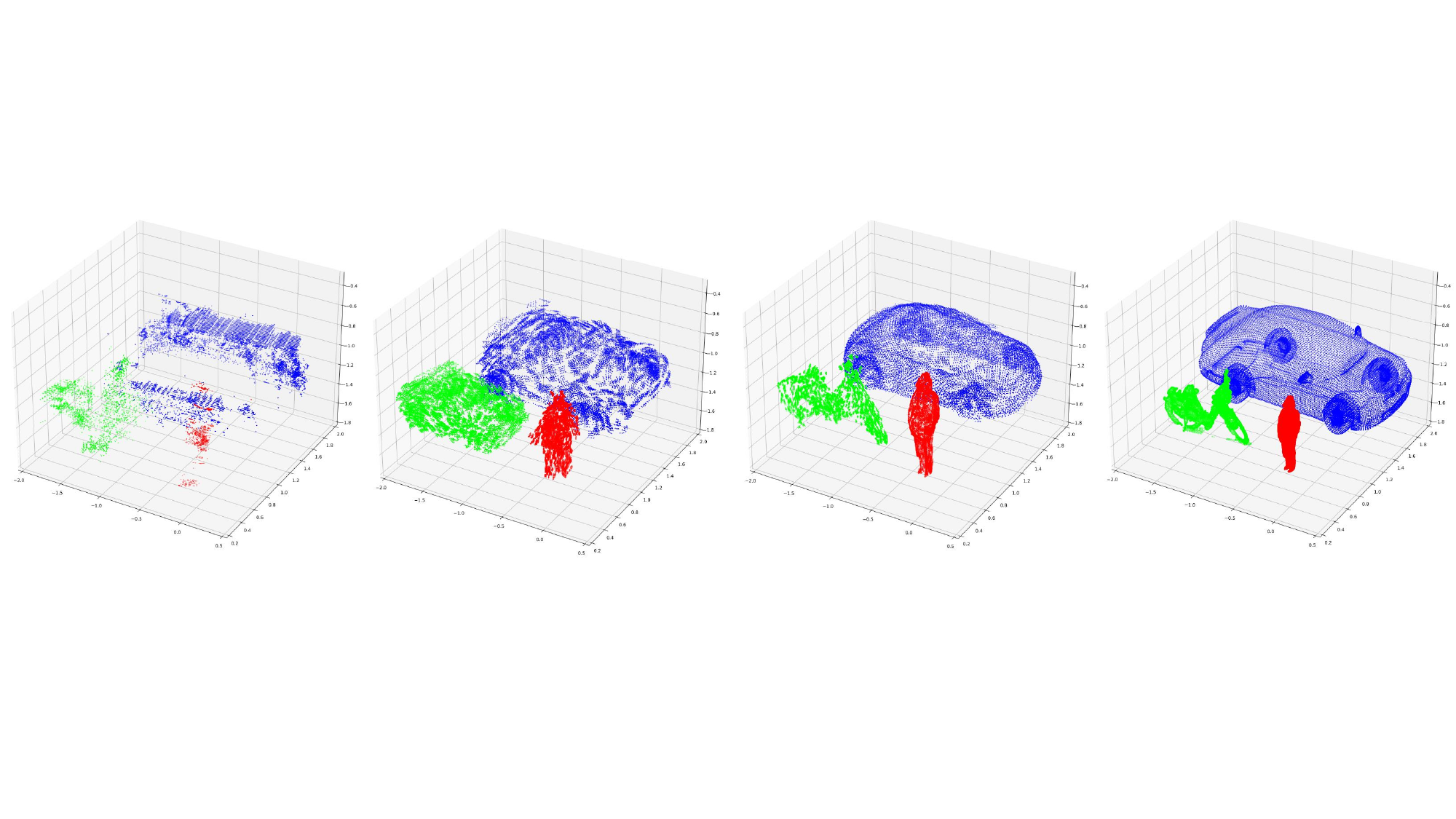}};
        \node at (-5.5, 3) {\small (a) \name's Radar Point Cloud};
        \node at (-1.5, 3) {\small (b) Baseline PCN};
        \node at (2, 3) {\small (c) \textbf{\name\ (Ours)}};
        \node at (5.8, 3) {\small (d) Ground Truth Shape};
    \end{tikzpicture}
    \vskip -1.2in
    \caption{mmWave Radar point clouds are very sparse due to their limited resolution and specularity of objects. In this paper, we first employ sensor fusion and odometry-aware temporal fusion to enhance mmWave radar point clouds, obtaining a denser point cloud (a). However, due to noise and artifacts, this point cloud is only a partial representation of objects. To reconstruct a complete representation, we leverage an encoder-decoder architecture to reconstruct complete shapes (c) from the enhanced mmWave points. The reconstructed point cloud closely resembles the ground truth objects in (d) and outperforms point completion baseline (b)  }
    \label{fig:teaser}
        \vskip -0.12in
\end{figure*}
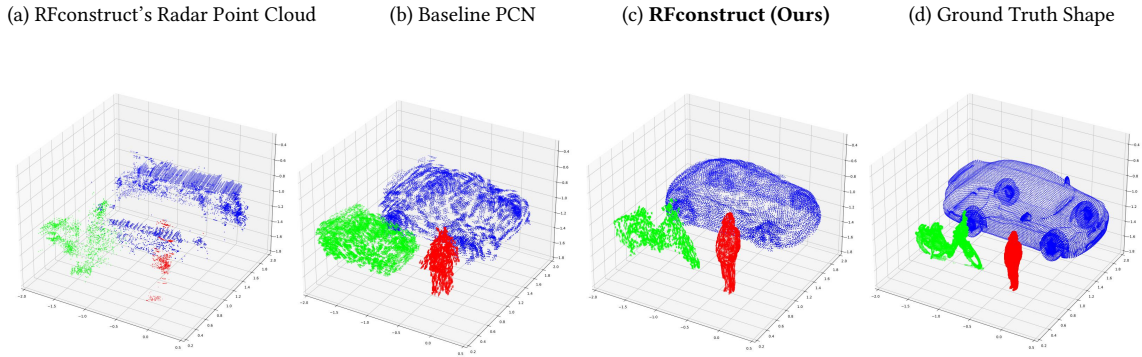

Enhancing mmWave radar point clouds has been studied in past work. \cite{akarsh2023radarhd} uses LiDAR supervision to generate denser point clouds from radar heatmaps. \cite{cai2023millipcd, lu2020millimap, lai2024panoradar} train neural networks to predict correct radar point clouds that match depth camera mappings of indoor environments. Other works focus on human pose estimation in controlled indoor scenarios\cite{lee2023hupr,zhao2018rfpose,zhao2018rfpose3d}. However, none of the prior work on mmWave radars can accurately reconstruct the 3D shape of common autonomous driving objects (cars, bikes, pedestrians) for partial radar measurements (see \S\ref{sec:related} for more related work discussion). 
On the other hand, 3D shape completion has been studied in the context of LiDAR and depth cameras~\cite{pcn, yu2023adapointr, qi2017pointnet, zhirong2015shapenets, wang2021voxel}. However, these works cannot be directly used for mmWave radars due to the unique nature of RF signals and the resolution limitations of the radars. First, radar point clouds are incredibly sparse compared to LiDAR point clouds as can be seen in Fig.~\ref{fig:teaser}(a). The angular resolution of radars can be $100\times$ lower than cameras and LiDARs on the azimuth plane and $2000\times$ lower along the elevation plane~\cite{Gao2021RAMP, ding2023hiddengems}.

Low resolution and lack of elevation information, however, are not the only limitations. Unlike light, mmWave signals do not scatter as much and mainly reflect off surfaces~\cite{lu2013measurement}. Hence, cars are highly specular and act as a mirror reflector of radar signals. As a result, most reflections never trace back to the mmWave receiver. This specularity makes certain portions of the car impossible to image, where a large portion of the car's surface is missing, as seen in Fig.~\ref{fig:teaser}(a). Finally, unlike vision, mmWave radar data is very scarce and highly dependent on the radar system that captures the data, making it very challenging to train deep neural networks. 

In this paper, we introduce \textit{\name}, the first system for 3D shape reconstruction from partial autonomous driving radar observations. It deploys COTS mmWave radars to generate interpretable 3D reconstructions of objects observed in autonomous driving scenarios, as shown Fig.~\ref{fig:teaser}(b). These scenarios result in brief, partial observations of objects in the scene. Our goal is first to obtain high-fidelity 3D point clouds from these observations using \name, then utilize a data-driven approach to generate complete 3D shapes of the partially observed objects. Enabling \name requires tackling domain-specific challenges including lack of high-resolution elevation information, sparsity of point clouds, signal specularity, and lack of training data, as explained below.

\vskip 0.1in \noindent $\bullet$ \textit{Single Dimension High Resolution: } 
        The angular resolution of radars depends on the number of antennas. Since the number of antennas becomes the bottleneck, some COTS mmWave radars trade-off elevation resolution to improve azimuth resolution~\cite{ticascade}. As a result, while a reasonably good azimuth resolution of $1.8^\circ$ could be achieved, the elevation resolution would be more than $10-20\times$ lower~\cite{ticascade}, making it nearly impossible to acquire dense, accurate detections in 3D. 
        To address this, instead of building a large 2D antenna array with $N\times N$ elements, \name\ leverages two orthogonally placed COTS radars to create two perpendicular long 1D antenna arrays with $N$ elements each. The horizontal and vertical arrays provide high resolutions along the azimuth and elevation dimensions but cannot resolve the ambiguity along the other orthogonal axis. \name\ fuses the data from the two arrays and correlates reflections to resolve the ambiguities and generate 3D radar point clouds from two 2D mmWave radars.

\vskip 0.1in \noindent $\bullet$ \textit{Specularity \& Sparsity: }  
Radar point clouds are sparse not only because of their low angular resolution. They also suffer from specularity and loss of information in the filtering process on the noisy raw radar heatmaps, as we elaborate in \S~\ref{sec:Background}. To address this, \name\ leverages the motion of the autonomous vehicle to illuminate and observe the same surface from multiple viewpoints. Even if the specular surface is not visible from many angles, it will reflect signals back to the radar as it moves to certain locations. \name\ accumulates reflections in many frames captured along the trajectory of the radar to compensate for missing reflections due to specularity in a single frame.
Moreover, observing objects from slightly shifted locations also provides us with more points. If the radar's per-frame displacement is smaller than its spatial resolution, accumulating frames helps obtain denser higher-resolution point clouds. However, frame-wise accumulation is nontrivial because the radar only estimates the location of objects relative to itself. As the radar moves and rotates over time, any errors in the location and orientation of the radar can cause smearing in the accumulated point clouds. \name\ first compensates for the ego-motion of the radar to obtain accurate absolute locations of the reflections.

\vskip 0.1in \noindent $\bullet$ \textit{Shape Reconstruction: } 
While \name's algorithms to fuse 2D radar data and accumulate it over time generate denser 3D radar point clouds, these point clouds continue to suffer from additional artifacts. First, they are still partial representations of objects, as viewpoints along the natural autonomous radar trajectory cannot cover all $360^\circ$ angles, i.e., there are still occluded parts.
Moreover, due to the multipath propagation of radar signals, some reflections bounce off the floor and other obstacles and trace back to the receiver, creating artifacts in the point clouds. Finally, imperfect tracking of the radar's ego-motion results in smearing in the point clouds. To address this, \name\ uses a point cloud completion deep neural network to reconstruct the full 3D shape. Although point cloud completion has been extensively studied in computer vision, employing existing techniques such as PCN~\cite{pcn} or AdaPoinTr~\cite{yu2023adapointr} cannot work on radar data, as we show in \S~\ref{sec:results}. Point cloud completion networks are typically designed and trained on partial but flawless point clouds without radar noise. This differs from radar point clouds, which typically contain considerable noise due to low resolution, leakage, and artifacts. They also assume prior knowledge of the bounding box of the shape being reconstructed, and their performance degrades significantly without it. \name\ builds on these works but modifies the architecture of the PCN network to allow it to learn to generate 3D shapes from noisy radar point clouds. Unlike prior work, \name\ does not require prior knowledge of the bounding box. Its performance degrades slightly without bounding box knowledge as opposed to with, as we show in \S~\ref{sec:results}

\vskip 0.1in \noindent $\bullet$ \textit{Scarcity of Radar Data: } 
Finally, collecting radar data is always a cumbersome and challenging task. While radar data sets are becoming more available~\cite{madani2022radatron, paek2022kradar, wang2021rodnet, zhang2023dual, Rebut_2022_CVPR}, none of the existing data sets can be used for \name. Most of these data sets are 2D or 3D with very low elevation resolution. The deployment scheme that \name\ uses to capture data, namely two orthogonal radars, varies significantly from the radars used in existing datasets. Moreover, these also do not have full 3D shapes of cars and bikes for ground truth. We address this by generating an augmented training dataset that consists of (A) Geometric Perturbation Augmentation of data generated from ShapeNet point clouds~\cite{zhirong2015shapenets}  and (B) simulated mmWave radar data generated using~\cite{guan2020hawkeye}. The combination of these two allows the network to learn the full shape of objects in the presence of noise and specularity that closely resembles radar data. While we train with synthetic and simulated data, we test \name\ with real data captured with our mmWave setup. Moreover, using our experimental setup, we collect a comprehensive dataset of objects that we use for evaluation and fine-tuning of \name. We show results with and without fine-tuning in~\S\ref{sec:finetune}.

To evaluate \name, we collect a real-world dataset using two 77-GHz TI MMWCAS mmWave radars~\cite{ticascade} on a TurtleBot3~\cite{turtlebot} and on a wheeled cart. We compare the 3D reconstruction performance of \name\ against a ground truth obtained from a full 3D scan of the objects using a depth camera and LiDAR-based reconstructions. Our dataset included cars, humans, bikes, and motorbikes. We use standard metrics like Chamfer Distance (CD) and Earth Mover Distance (EMD) for our evaluation. We show that \name\ can generate accurate point cloud reconstructions outperforming baselines. Fig.~\ref{fig:teaser} shows a qualitative example of \name's results. Our quantitative results reveal the following: without bounding box priors, \name\ achieves a mean CD of 15.3 cm compared to 42.1 cm for PCN~\cite{pcn} and 43 cm for AdaPointTr~\cite{yu2023adapointr}. \name\ also achieves a mean EMD of 49.9 cm compared to 90.0cm for PCN and 91.46 cm for AdaPointTr. This shows that, unlike past work, \name\ does not necessarily require knowledge of the bounding boxes.  With bounding box priors, \name\ achieves a mean CD of 2.01 cm compared to 4.58 cm for PCN and 4.10 cm for AdaPointTr. \name\ also achieves a mean EMD of 6.25 cm compared to 13.03cm for PCN and 17.5 cm for AdaPointTr. Moreover, we show that \name's temporal fusion significantly outperforms SAR since SAR is extremely sensitive to a single millimeter error in the positions of the antennas. We also present extensive microbenchmarks, ablation studies, and qualitative results in~\S\ref{sec:results}.

This paper makes the following contributions:

\begin{itemize} [leftmargin=*,topsep=3pt]

    \item \name, to the best of our knowledge, is the \emph{first} mmWave radar system capable of reconstructing 3D shapes and details of commonly found street-side objects  (cars, pedestrians, bikes and motorbikes, ) from partial observations of driving past the object.
   
    \item We present a novel system design that combines: (1) Radar fusion from two orthogonally placed radar devices to simultaneously achieve high azimuth and elevation resolutions, (2)  Odometry-aware temporal fusion captured along the motion trajectory to combat specularity and sparsity in the point clouds, and (3) a radar shape completion deep neural network to reconstruct complete 3D shapes from noisy and incomplete point clouds. 

    \item We provide a new radar dataset composed of: (1) an augmented training dataset generated from ShapeNet-55 \cite{shapnet} to emulate mmWave radar point cloud imperfections such as specularity and noise, (2) a simulated radar dataset that captures radar characteristics like sinc leakage, (3)  a real 3D mmWave radar dataset with equally good resolution in azimuth and elevation collected on our prototype radar. This dataset contains more than 100,000 raw radar frames with 162 cars, 91 bikes, and 52 pedestrian data points that are paired with depth camera recordings and odometry information.  

    \item We build, implement, and extensively evaluate a prototype of \name\ in real settings using COTS MIMO cascaded radars.

\end{itemize}

\vskip 0.06in \noindent {\bf Limitations: } While taking the first steps toward 3D shape reconstruction from mmWave radars in autonomous driving, \name has several limitations that need to be addressed before we can have a fully functional system. Most notably, our current implementation of \name only works for static objects. Moving objects requires further research as it significantly complicates temporal fusion which will depend on the relative speed and positioning of the cars and can lead to smearing. Moreover, we only trained and tested \name for three classes of objects which we deem as the most important starting point. To make \name more general, we need to extend it to new classes like trees, road signs, fire hydrants, trash cans, large trucks, etc. 
We discuss these and additional limitations of \name and how to address them in more detail in~\S\ref{sec:limitations}.

\section{Related Work}
\label{sec:related}

\vskip 0.1in \noindent {\bf  A. Millimeter-wave Radar Perception:}
Recent years have witnessed an increasing interest in mmWave radar perception for various applications, ranging from human posture tracking~\cite{zhao2018rfpose, kong2022m3track, lee2023hupr} to gesture recognition~\cite{lien2016soli, li2023gesture}. On autonomous vehicles and autonomous mobile robots, mmWave radars have also been exploited for odometry~\cite{lu2020milliego,Almalioglu2021millirio}, mapping~\cite{lu2020millimap, chen2022floorplan, akarsh2023radarhd}, object detection~\cite{Danzer2019PointNets, major2019Qualcomm, bansal2020pointillism, dong2020probabilistic, Ouaknine2021ICCV, Gao2021RAMP, meyer2021graph, zhang2021raddet, madani2022radatron, paek2022kradar, liu2024smurf, imwutbbox}, and scene flow estimation\cite{ding2023hiddengems}.
These works only output high-level abstract features for specific end goals, such as semantic segmentation of radar point clouds~\cite{Ole2017LSTM,Ole2018segmentation, Ouaknine2021ICCV}, coordinates of objects~\cite{Gao2021RAMP}, 2D BEV bounding boxes~\cite{Danzer2019PointNets, major2019Qualcomm, dong2020probabilistic, zhang2021raddet, madani2022radatron}, and 3D bounding boxes~\cite{bansal2020pointillism, meyer2021graph, paek2022kradar, liu2024smurf}. In contrast, \name\ aims to reconstruct 3D shapes of objects, which contain  high-frequency details and contextual and perceptual information than the abstracted features.

\vskip 0.1in \noindent {\bf B. 3D Reconstruction with mmWave Radars:} 
There are prior works that try to reconstruct 3D shapes, but they focus on reconstructing 3D meshes of human bodies~\cite{zhao2019rfavatar,xue2021mmmesh,chen2022mmbody,xie2023mmPoint}. They are specifically designed and trained for human targets and cannot generalize to other types of objects. \cite{guan2020hawkeye, lai2024panoradar} create 2D depth maps of cars and indoor buildings from radar heatmaps. However, when converting depth maps to 3D point clouds, they suffer from common inaccuracies and artifacts and are incomplete because of occlusion. Sun et al.~\cite{sun20213drimr, sun2022r2p, sun2022multiple} take a step further, combining multiple GAN-generated depth maps from multiple views to generate a complete point cloud and feeding it to another generative model to reconstruct 3D shapes. These methods, however, have only been shown to work on cars and struggle with overfitting and insufficient input information due to requiring two stages of deep learning (Radar to depth map and depth map to 3D shape).

\vskip 0.1in \noindent {\bf C. Coherent and non-coherent radar fusion:}
Coherent combination methods such as Synthetic Aperture Radar (SAR) and inverse SAR~\cite{akarsh2020osprey} are used to improve the angular resolution of radar imaging. SAR emulates a much larger virtual antenna aperture by mechanically scanning a single radar device to cover the aperture. 
Prior work~\cite{guan2020hawkeye, yanik2020mimosar} uses horizontal and vertical sliders to move the radar to emulate a planar array. However, this method is limited by its long scanning time and cumbersome setup. Instead of introducing mechanical scanning on slide rails, other work leverages the movement of autonomous systems to emulate a larger aperture~\cite{qian2020millipoint}. However, these methods are highly vulnerable to millimeter-level errors in the antenna locations across motion trajectory as well as any slight rotation in the array due to motion. Moreover, they can only improve the resolution along the dimension of the moving trajectory of the robot and cannot reconstruct shapes from the point clouds. In contrast, \name\ can also improve the elevation angle resolution, which is perpendicular to the vehicle trajectory. \name\ also constructs complete shapes from partial observations and does not require mm-level localization accuracy.

Some prior work implicitly leverages consecutive radar frames using neural networks: Millimap~\cite{lu2020millimap} feeds multiple BEV heatmaps in consecutive frames into a neural network to generates to generate indoor floor maps without specularity artifacts; \cite{zhao2018rfpose,zhao2018rfpose3d} leverages time series of radar frames to estimate human skeleton; \cite{li2022temporal} uses attention layers to learn the relations between successive BEV radar heatmaps for better object detection. MilliPCD\cite{cai2023millipcd} incoherently combines patches of hand-held radar scans to generate a coarse point cloud of indoor spaces. It then uses a noise-aware shape re-constructor model to generate a fine-grained point cloud from the coarse, noisy input. These works, however, target indoor mapping scenarios such as hallways.

\vskip 0.1in \noindent {\bf D. Multi-Radar combination:} 
Some MIMO radars do use perpendicular TX and RX arrays to emulate a 2D rectangular virtual antenna array~\cite{wang2008orthogonal, Shirakawa20133DScanMR}. However, more than 100 physical TX and RX antennas are needed to achieve sub-degree angular resolution, which is beyond the scalability of today's MIMO radar hardware. Therefore, commercial MIMO radars nowadays tend trade off elevation resolution to have higher azimuth angular resolution~\cite{ticascade}.
\name\ uses two orthogonally placed commercial MIMO radars and leverages post-processing to achieve high azimuth and elevation resolutions at the same time. Some prior works tried similar placement of single-chip MIMO radars~\cite{wang2021rodnet, adhikari2022mishape, wu2023rfmask}, as well as cascaded MIMO radars~\cite{rahman2025mmvr, icramultirad}. However, they do not explicitly leverage the two arrays to boost the angular resolution and enhance the radar point cloud. Instead, they directly feed ADC samples~\cite{adhikari2022mishape} or heatmaps from each radar~\cite{wu2023rfmask, lee2023hupr, rahman2025mmvr} to a neural network to directly perform tasks like human pose estimation~\cite{rahman2025mmvr, lee2023hupr}, human silhouette segmentation~\cite{adhikari2022mishape, wu2023rfmask}, odometry estimation~\cite{icramultirad}, or object detection in BEV~\cite{wang2021rodnet}. In contrast, \name\ combines the high-resolution dimension of the two radars explicitly to enhance radar point clouds, which it then uses for shape reconstruction.

\vskip 0.1in \noindent {\bf E. Point-Cloud Completion:} 
Learning-based methods have become the current research trend in point cloud completion, and they can be further categorized into voxel-based and point-based methods. The voxel-based method entails the voxelization of the disordered point cloud, followed by the use of the voxelized 3-D model to accomplish shape completion~\cite{zhirong2015shapenets, wang2021voxel}. Point feature-based approaches leverage deep neural network architectures designed to work directly with point clouds like PointNet~\cite{qi2017pointnet}. \name\ builds on top of PCN~\cite{pcn}, which has an encoder-decoder architecture and uses the strategy of coarse-to-fine point generation. The encoder extracts global features from the partial input, and the decoder uses these global features to generate a dense, complete point cloud.

\section{Background and challenges of Millimeter-Wave Radar Imaging }
\label{sec:Background}

\begin{figure*}[t!]
    \begin{center}
        \includegraphics[width=\textwidth]{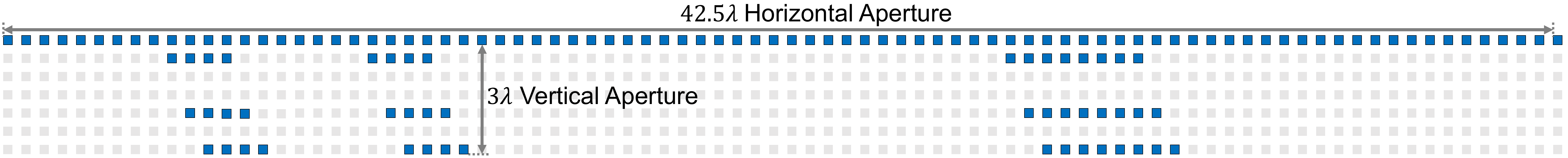}
        \caption{Virtual Antenna Array Topology of COTS mmWave Radar: \textnormal{Long 2D array ($42.5\lambda \times 3\lambda$ aperture) with high azimuth resolution and low elevation resolution.}}
        \label{fig:virtual}
    \end{center}
\end{figure*}

\subsection{3D Radar Imaging} 
Millimeter-wave radars image the environment by transmitting radar waveforms to the 3D space and estimating reflected signal power in a 3D spherical coordinates defined by range ($\rho$), azimuth angle ($\phi$) and elevation angle ($\theta$). The range and angular resolutions of the 3D radar heatmaps are determined by the signal bandwidth and the size of the antenna array, respectively. 

A 2D antenna array is needed to localize reflectors in the 3D spherical coordinates $(r, \phi, \theta)$, as the direction of reflector results in different phases at different antennas in the 2D array:

\begin{equation}
\label{eq:channel_2d}
S_{m,k}(r) \propto e^{-j\frac{2\pi}{\lambda}(md sin\theta cos\phi + kd cos\theta)} 
\end{equation}
where $\lambda$ is the wavelength, $d =\lambda/2$ is the separation
between consecutive elements, and $S_{m,k}(r)$ is the signal reflected by an object at distance r and then received on 
the antenna element index by $(m,k)$ in the 2D array.
Therefore, one can compute the reflected power $x$ along the spherical angles $\phi$ and $\theta$ by
adding a phase shift to the signal received on every antenna before
combining the signals~\cite{ArrayDSP}:

\begin{equation}
x(r, \theta, \phi) = \sum_{m=0}^N\sum_{k=0}^N{S_{m,k}(r)e^{j\frac{2\pi}{\lambda} [m d
sin\theta cos\phi + k d cos\theta]}}
\end{equation}

\subsection{Resolution of Imaging Radars.} 
Although most mmWave radars today can achieve good depth resolutions (4-10 cm), their angular resolutions are nowhere near those of cameras and LiDARs. This is because unlike cameras or LiDARs, radars cannot directly distinguish lights coming from different angles using lenses or narrow laser beams. Instead, radars use an array of antennas that transmit and receive signals from all angles. Reflections from different angles are then resolved from the superposition by estimating the phase differences of the received signals across the antenna array, which is proportional to $cos\theta$, where $\theta$ is the incident angle. 

However, angle estimation results in an ambiguity function as a {\it sinc} function. The width of the main-lobe of the {\it sinc} function is inversely proportional to the size of the antenna array, as is the angular resolution.
The angular resolution of an antenna array with $N$ elements separated by half wavelength ($\frac{\lambda}{2}$) can be approximated by:
\textit{Angular Resolution} $\approx \frac{2}{N sin\theta}$.

As a result, a point reflector in the scene is convolved with very wide {\it sinc} functions along both the azimuth and elevation axes in the resulting radar heatmaps, which eliminates all high-frequency shapes and details in the radar heatmaps. This is because, mathematically, convolution with a {\it sinc} is equivalent to multiplying the frequency domain with a rectangle function, effectively zeroing out high frequencies~\cite{sayeed2010continuous}. 
The side-lobes of the {\it sinc} function also create noise in the heatmap.

To achieve sufficient resolution in range, azimuth, and elevation, a mmWave radar requires a large 2D antenna array, which introduces significant hardware complexity. In practice, most COTS mmWave radars can form a large virtual antenna array along only one dimension, while the second dimension typically has a much smaller aperture and sparser antenna elements, often spaced beyond $\lambda/2$, as shown in figure \ref{fig:virtual}. This leads to poor angular resolution and the presence of grating lobes, making accurate target detection along the second dimension highly challenging.

\subsection{Radar Point Cloud Processing.} 
The high-dimensional volumetric representation of 3D radar heatmaps are very sparse and noisy, yet comes with a very large volume, which puts a big burden on the data transfer and storage.
Therefore, it is common practice in commercial mmWave radar devices to compress 3D radar heatmaps into a point-cloud format.  
The extraction of point clouds from heatmaps is done through a threshold process.
For every voxel in the radar heatmap, if the corresponding reflection power is above a power threshold, the voxel is mapped to a point in the point cloud.

The challenge is to determine the power threshold, because
a too high threshold leads to missed detections and loss of useful information, 
while a too low threshold results in very noisy point clouds that contain false alarms due to noise, side lobes of the 2D {\it sinc} function, and background clutter. The Constant False Alarm Rate (CFAR) algorithm and its variants such as Cell Averaging CFAR (CA-CFAR) and Ordered Statistic CFAR (OS-CFAR) are commonly used to convert radar heatmaps into point clouds. 
CFAR is an adaptive detection algorithm that adjusts the detection threshold in accordance with the measured background. Specifically, a CFAR detector estimates the noise floor for the cell under test by analyzing data from neighboring cells.

\section{\name}
\label{sec:method}

\begin{figure*}[t!]
\begin{center}
\includegraphics[width=16cm]{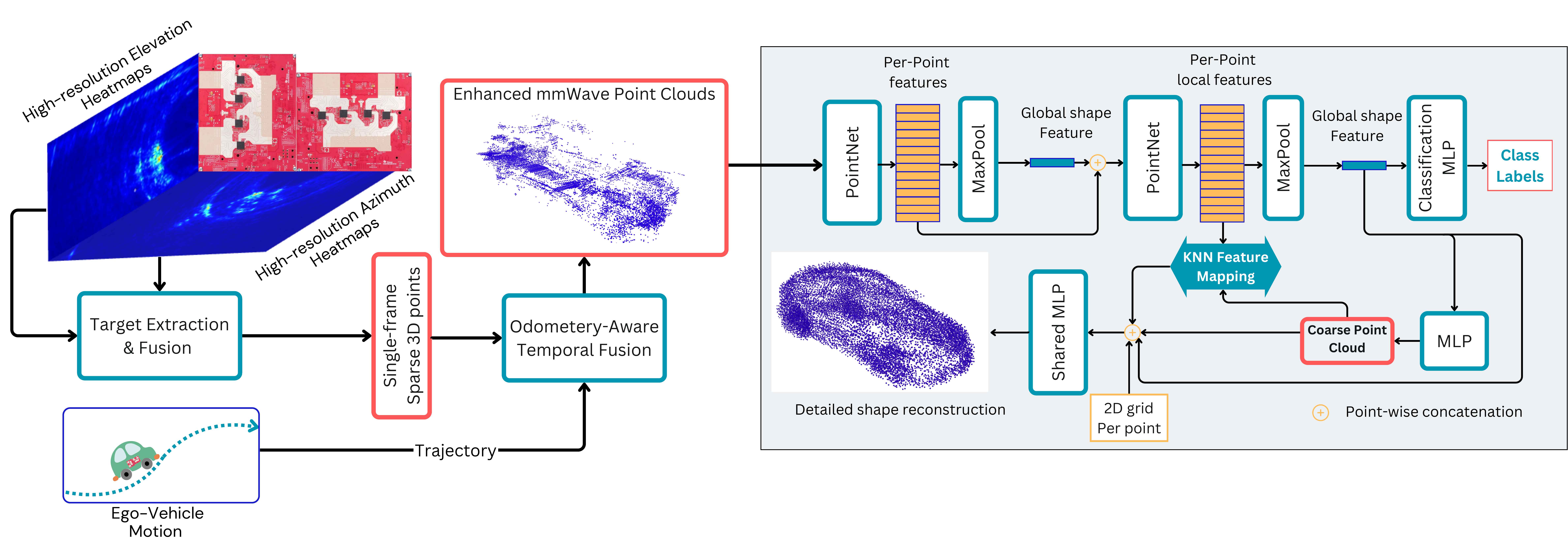}
\vskip 0.0in \caption{\name\ Overview. In Stage 1 (left), Enhanced, dense point clouds are generated by multi-radar fusion and odometry-aware fusion. In stage 2 (right), \name's completion network generates complete 3D shapes from the enhanced point clouds. 
}
\label{fig:overview}
\end{center}
 \vskip -0.2in
\end{figure*}

\name\ is a 3D reconstruction framework designed for mmWave radars, which can recover complete 3D shapes of objects commonly seen by self-driving cars such as vehicles, motorcycles, bicycles and pedestrians. 
\name\ operates in two stages shown in Fig.~\ref{fig:overview}; The first stage focuses on the enhancement of mmWave radar point clouds. The second stage leverages the enhanced partial point cloud to feed through a shape-completion network that generates a complete 3D shape from partial inputs.

\subsection{Enhanced mmWave Radar System}
\label{subsec:radar}

\subsubsection{Orthogonal Radars Fusion: Enhancing elevation resolution} 
The first and foremost challenge \name\ tries to overcome is the low angular resolutions of COTS mmWave radars. 
As explained in $\S$~\ref{sec:Background}, the angular resolution of a radar is bounded by the number of antennas.
High-resolution 3D radar imaging requires a large 2D antenna array. According to \S~\ref{sec:Background}, $N\times N=N^2=10000$ antenna channels are needed to achieve $\approx1^\circ$ angular resolution.
Although MIMO radar~\cite{sun2020mimo} can emulate a 2D virtual antenna array with $N\times N$ elements using $N$ transmitters (TX) and $N$ receivers (RX), $N=100$ physical TX and RX channels are still orders of magnitude more than those in today's MIMO radar systems.
To overcome this challenge, we try to enable equally high-resolution 3D imaging using two 1D MIMO radars with $\sqrt{N}$ TX and $\sqrt{N}$ RX.

\begin{figure}[t!]
    \centering
    \begin{minipage}[b]{0.4\textwidth}
        \centering
        \includegraphics[width=\textwidth]{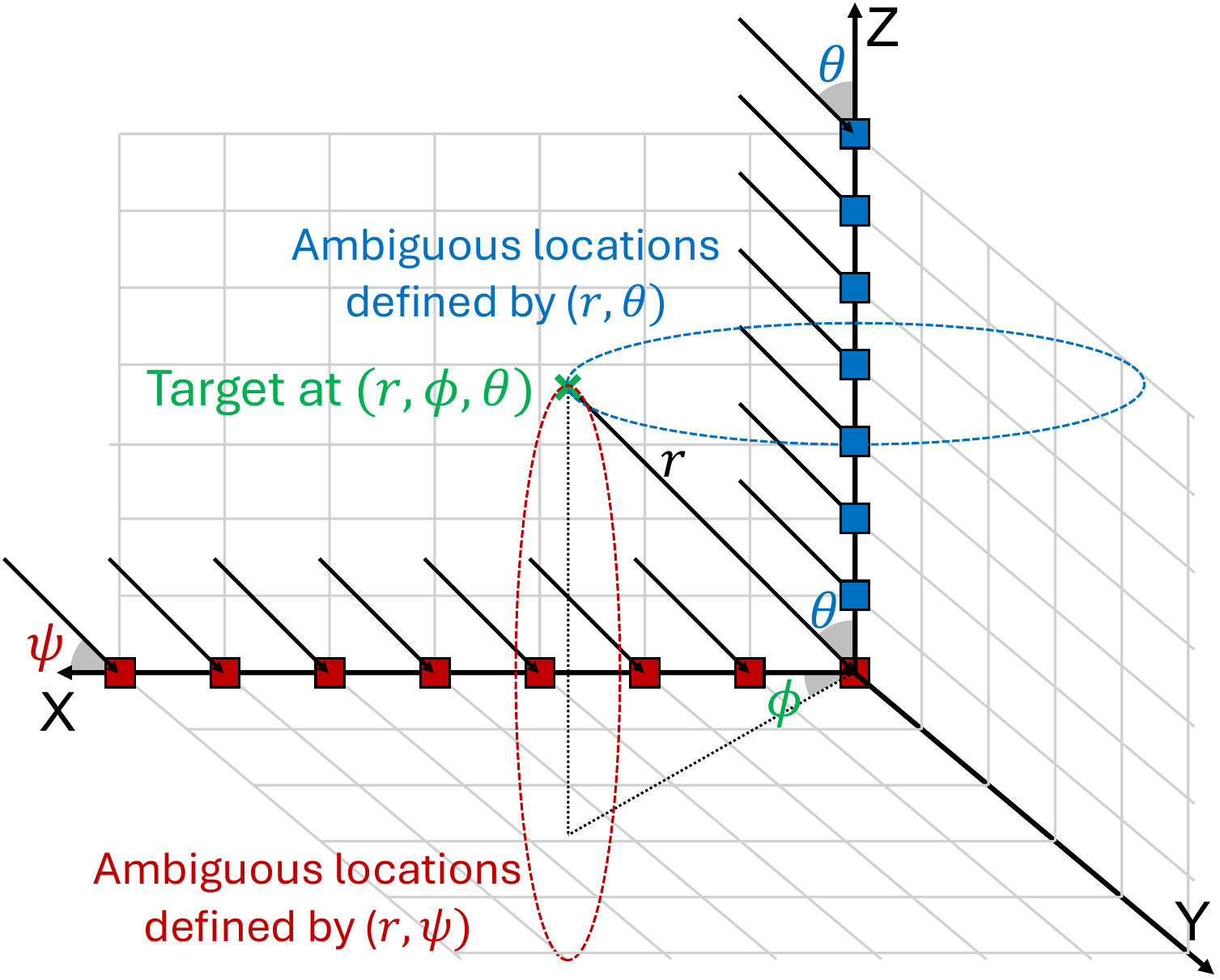}
        \caption{Perpendicular 1D virtual arrays in 3D space.}
        \label{fig:spherical_coord}
    \end{minipage}
    \hfill
    \begin{minipage}[b]{0.54\textwidth}
        \centering
        \includegraphics[width=\textwidth]{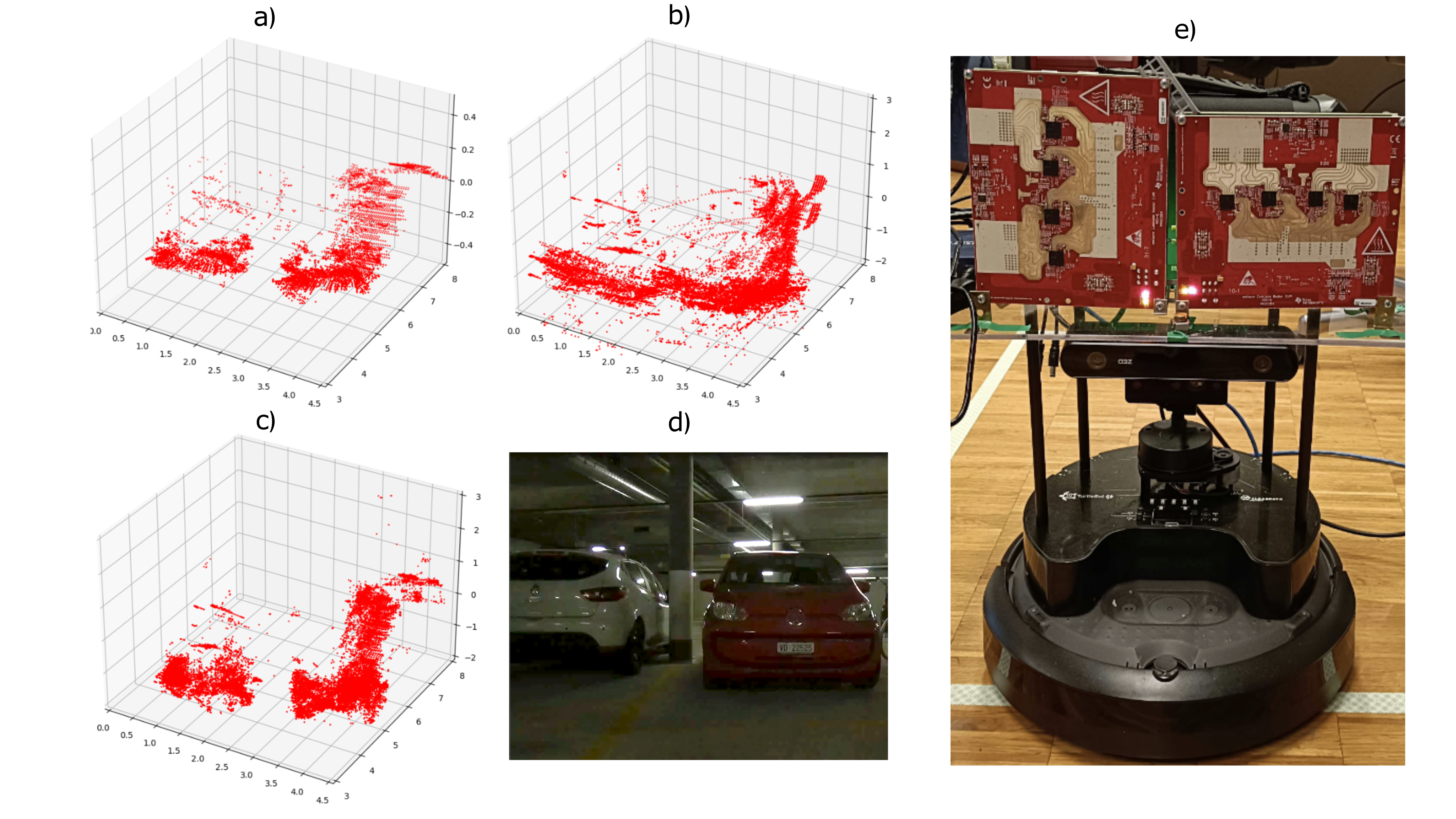}
        \caption{Temporal fusion for (a) horizontal radar only, (b) vertical radar only, and (c) the combined vertical and horizontal output. (d) is a camera picture of the scene. (e) is our experimental setup.}
        \label{fig:hv_fusion}
    \end{minipage}

\end{figure}

Imaging a 3D scene using a 1D antenna array results in ambiguities on the other dimension.
However, radar reflections in the 3D space are also very sparse, so we try to leverage this sparsity to resolve the ambiguities.
We jointly leverage two perpendicularly placed 1D antenna arrays as shown in Fig.~\ref{fig:spherical_coord}.
The intuition is that, although it suffers from elevation ambiguities, the horizontal radar $R1$ can emulate a 1D horizontal antenna array with $\sqrt{N} \times \sqrt{N}=N$ elements, and achieve an equally high azimuth resolution as a 2D $N\times N$ MIMO radar. 
Similarly, the vertical radar $R2$ has a high elevation resolution but cannot estimate the azimuth angle. 
For a target detected at the azimuth angle $\phi_t$ with $R1$, if we could estimate its elevation angle using $R2$, we would be able to get the best of both worlds and enable high-resolution 3D imaging.
To understand whether combining the resolution of perpendicularly placed 1D radars is possible, we first need to understand the ambiguity that must be resolved.
When imaging a 3D scene using the 1D vertical array of $R2$, all reflections coming from the circular ring area defined by the range $r$ and the elevation angle $\theta$ (as shown in Fig.~\ref{fig:spherical_coord}) collapse into the same bin in the 2D range-elevation heatmap, and we are unable to resolve their azimuth angle $\phi$.
On the other hand, in the range-azimuth heatmap measured by horizontal radar $R1$, all reflections coming from the circular ring area (as shown in Fig.~\ref{fig:spherical_coord}) defined by $r$ and the angle between the direction of the target and the x axis (represented by $\psi$) collapse to the same bin. Note that $\psi \neq \phi$, but $\phi$ can be computed from $\psi$ as $cos\psi = sin\theta cos\phi$.

The key to solving the ambiguities is to associate the reflections detected by both radars.
First, we ensure that the two radars have the same range resolution, so that the same target is segmented into the same range bin in the two radars. 
If there is only one reflector in each range bin, we can directly pair its elevation angle $\theta$ estimated in $R2$ with azimuth angle $\phi=arccos(cos\psi/sin\theta)$, and localize the reflector in the 3D.
However, when multiple reflectors are in the same range bin, we must find a one-to-one mapping between the two radars.

\subsubsection{Orthogonal Radars Fusion: Target Detection \& Association}

Finding corresponding targets in multiple radar heatmaps is challenging, because radar heatmaps lack rich features as in RGB camera images that can be used to associate pixels in multiple camera images~\cite{yang2022distributed}.
Radar heatmaps appear as the reflected signal power, which is the key feature we can leverage to identify and associate targets.

For simplicity, let us first assume that although multiple reflectors are in the same range bin, they are resolvable in both radars because all reflectors fall into different azimuth and elevation bins. In this case, the numbers of targets that we should detect in each radar are the same and we can find a one-to-one mapping for all targets. For the detected targets, we put them in the order of signal power and then associate targets with the closest power level in the two radars. However, if two targets fall into the same azimuth bin in the horizontal radar $R1$ or in the same elevation bin in the vertical radar $R2$, they become indistinguishable. As a result, there will be a discrepancy in the number of targets detected, and finding a one-to-one mapping becomes implausible.

Note that COTS MIMO radars are not exactly 1D. As can be seen in Fig.~\ref{fig:virtual}, while most of the virtual antennas sit in the same row, forming a uniform 1D array, there are some elements located in other rows. It is plausible to further resolve different targets along the elevation axis leveraging these vertical antennas. However, they provide us with only $19^\circ$ elevation resolution. In addition, the 2D virtual array is very sparse, which results in severe grating lobes that prevent us from accurately detecting the targets.

Without an effective way to further resolve ambiguities in the 1D radar heatmaps, when we suspect ambiguities, i.e., when the numbers of targets detected in the two radars do not match, we have to take a more conservative approach towards associating targets. We first list the targets detected by each radar in the order of the reflected signal power. Then we start associating targets from the highest power to the lowest until the power discrepancy becomes larger than 3dB. Although this target association scheme would throw away a lot of targets, making the point cloud even sparser, it avoids placing points at wrong locations. Besides, we can rely on the next module in the radar signal processing pipeline, temporal fusion, to compensate for the sparsity. We found experimentally that this approach outperforms creating denser but noisier point clouds with spurious points due to incorrect association.

It is also worth mentioning that the field of views (FoV) of the two radars in the azimuth and elevation dimensions are different because the radiation patterns of the patch antennas of MIMO radars are not quasi-omnidirectional. 
According to the antenna beam patterns shown in~\cite{timmwcas_userguide}, while the azimuth FoV for $R1$ is $\pm 70^\circ$, its elevation FoV is only $\pm 20^\circ$. On the other hand, $R2$ has $\pm 20^\circ$ azimuth FoV and $\pm 70^\circ$ elevation FoV.
However, we must ensure that the target we try to detect is visible to both radars, so we limit the FoV of the combined radar system to $\pm 45^\circ$ in both azimuth and elevation dimensions. 
Moreover, we scale the signal power in different angular bins to compensate for the different antenna gains in different directions in order to correctly associate targets.
After mapping targets from planar heatmaps to 3D space, we convert their spherical coordinates to cartesian coordinates for further processing. 
We also filter out points below the ground or higher than 3m, as they do not represent any practical targets.

\subsubsection{Odometry-Aware Temporal Fusion}
Although combining the measurements of two radars allows us to achieve higher resolution, the limited number of targets that we can resolve makes the super-resolution 3D radar point clouds even sparser. To capture dense point clouds of objects, \name\ leverages the radar motion along with the self-driving car, so it can observe the same objects from different view points along the trajectory.
\name\ thus combines a number of consecutive radar frames and accumulates their point clouds to obtain denser point clouds.

Radar point clouds in different frames cannot be directly concatenated because they measure the relative position of the reflectors to the radar system. To accurately accumulate point clouds in different frames, we first need to convert the relative reflectors' positions into absolute positions by compensating for the ego-motion of the radar. 
To obtain the position and orientation of the radar per frame, \name\ uses an odometry sensor that is synchronized with the radar.

For each frame of 3D point clouds, we first obtain the corresponding position and orientation from the odometry.
We then apply the necessary translation and rotation to adjust the point clouds to the same origin, which is the first reference frame. 
This is done for all frames through the experiment trajectory. 
Then, the transformed point clouds are merged to generate a denser 3D point cloud. 

The accuracy of the accumulated point clouds depends on the accuracy of the position and orientation tracking of the radar. The longer the duration and trajectory over which we accumulate, the larger the positioning error and more smearing the resulting point cloud will have. 
Therefore, we limit the temporal fusion to a short trajectory with an average length of about $3 meters$, with a radar frame rate of approximately $60 fps$. 
In \ref{sec:results}, we show the effect of increasing or decreasing the trajectory length and number of frames on the 3D reconstruction.

The temporal fusion provides us with denser point clouds. However, due to the limited trajectory in a practical setting, the resulting point clouds still suffer from some specularity and missing regions.
This limitation motivated us in the direction of point cloud completion, where \name\ combines its enhanced radar point clouds with learned point cloud completion methods to generate 3D shapes.

\subsection{Shape Completion from Radar Point Clouds}
\name\ takes a data-driven approach to reconstruct complete 3D shapes with missing parts filled in from partial and still sparse radar point clouds. 
To better leverage geometric priors of the 3D shape to be reconstructed, we focus on four types of object that are most commonly seen and are most important for self-driving cars. They are cars, bicycles, motorcycles, and pedestrians.

\subsubsection{\name's Completion Network}

We take advantage of the recent advances in point cloud completion in the computer vision domain. In particular, we primarily conform and adapt PCN~\cite{pcn} to the completion of radar point clouds. However, adapting PCN to the radar domain is challenging for the following reasons. 

First of all, PCN is trained with artificial point cloud data, generated from perfect 3D shapes through downsampling and cropping, which is very different from real-world radar point clouds.
Therefore, \name's network needs to learn the 3D shapes of objects from the much sparser, specular, and noisier real-world radar data. 

Moreover, The original PCN makes many restrictive assumptions about the input partial point clouds. The partial input needs to be normalized with prior knowledge about the complete input dimensions, such that the full shape of the object lies within a unit sphere. In addition, the full shape of the object needs to be centered at the origin and placed in a specific orientation. These assumptions are not practical in a real-world setting unless we poses the knowledge of all the objects' bounding boxes. We are also interested in achieving an end-to-end system that can directly generate 3D shapes without the assistance of another bounding-box estimation network. 

\name\ follows an encoder-decoder architecture similar to that of PCN. However, we introduce two main architectural changes that aid the encoder in learning more refined features and injecting local point features in the decoder shape generation. \name's neural network architecture is shown in Fig.~\ref{fig:overview} and detailed below. 

\vskip 0.05in
\noindent \textbf{Encoder.}
The encoder network of \name\ takes N partial input points that are represented as $N \times 3$ matrix $X$, where each element $x_i$ of this matrix consists of the $(x, y, z)$ coordinates of the point. The encoder consists of two cascaded PointNet (PN)~\cite{Danzer2019PointNets} layers. Each PN layer consists of a shared MLP followed by a Point-wise Maxpool. The first layer takes the matrix $X$ as input, and first produces point features $F_1$ from the shared MLP. These point features are then combined into a global feature vector $g_1$ via the maxpooling layer. The second PN layer input is a per-point concatenation of the point features $F_1$ and global feature vector $g_1$, which represents an augmented point feature matrix. The output of the second layer is local point features $F_2$ and a global feature vector $g_2$.

\vskip 0.05in
\noindent \textbf{Classification-Driven Feature Refinement.}
A classifier module is responsible for classifying the input objects based on the features produced by the encoder. The output object classes of the classifier module are not directly used in our system but are used in the computation of the loss function. The intuition behind this is the classifier guiding the encoder to better differentiate the four types of targets and produce more class-differentiable features. This enables the encoder to still be able to produce robust features from the noisy and misaligned partial inputs provided by the mmWave radars. The classifier module consists of an MLP layer, and its input is the final global feature vector of the encoder $g_2$, and its output is the class label $cl$. Note that the classifier module is only used in training.

\vskip 0.05in
\noindent \textbf{Local-Global Context Modeling Decoder.}
In the original PCN, the only input passed between the encoder and the decoder was a global feature vector. This means that the generated shape from the decoder uses very high-level features and can miss or discard important features that would be considered at the point level. This is particularly harmful to radar data because the nature of radar noise can lead to the generation of a very rough feature vector. 

In our architecture, we rectify this by including the point features in the decoder input. The decoder is a two-stage decoder, the first stage generates a coarse point cloud that constitutes the general object shape, and the second stage generates a fine-grained point cloud from the coarse input. The first stage is an MLP module whose input is the global feature vector $g_2$ generated by the encoder. The output of this MLP is reshaped to $(n, 3)$ where $n$ is the number of coarse points. 

At this point, we measure a nearest-neighbor correspondence between the generated coarse point cloud and the partial input point cloud. This correspondence is used to extract local point feature vectors from $F_2$ encoder vector. These local point features are concatenated to the coarse output as well as the global feature vector to generate a hybrid local-global feature embedding per point in the coarse point cloud.

A patch of $u^2$ points is generated around every coarse point in the coarse point cloud. The concatenated matrix is then passed through a shared MLP that deforms the added patch of points based on the global and local features and generates the fine-grained output.

\subsubsection{Loss Functions}

The loss for our network is a combination of three weighted losses, classification loss, coarse loss, and fine loss. The classification loss is a softmax cross-entropy loss that measures the loss between the predicted class and the true class. For a batch of \( N \) examples, the classification loss is the average of the cross-entropy loss over the $N$ examples:
\begin{equation}
L_{\text{class}} = -\frac{1}{N} \sum_{i=1}^N \sum_{j=1}^K y_{ij} \log(\hat{y}_{ij})
\label{eq:classloss}
\end{equation}
where \( y_{ij} \) is the true label and \( \hat{y}_{ij} \) is the predicted probability for class \( j \) of example \( i \).

The coarse loss measures the loss between the output of the coarse stage in the decoder and the ground truth using the Earth Mover's Distance (EMD) metric~\cite{fan2017emd}. EMD finds a bijection $\phi: A \rightarrow B$ which minimizes the average distance between closest points. The optimal $\phi$ is computationally expensive to fine, which is why we implement an iterative approximation method to measure it. Moreover, we use EMD to compute the coarse loss the number of coarse points is smaller than the number of fine points, which accelerates the training process.
\begin{equation}
    EMD(A,B) = \min_{\phi: A\rightarrow B}  \frac{1}{|A|}  \sum_{a \in A} \|a - \phi(a) \|^2
\end{equation}

Meanwhile, the fine loss is measured between the final output and the ground truth using Chamfer Distance (CD)~\cite{fan2017emd} as the metric. CD measures the average bidirectional closest-point distance between the output and ground-truth point clouds. 
The Chamfer distance between two point sets \( A \) and \( B \) is given by:
\begin{equation}
CD(A, B) = \frac{1}{|A|} \sum_{a \in A} \min_{b \in B} \|a - b\|^2 + \frac{1}{|B|} \sum_{b \in B} \min_{a \in A} \|b - a\|^2
\label{eq:chamfer}
\end{equation}

Our combined loss function is a weighted sum of the three aforementioned components: 

        $$L = L_{\text{EMD}}(A_{\text{coarse}}, A_{\text{gt}}) + \alpha L_{\text{CD}}(A_{\text{fine}}, A_{\text{gt}}) 
        + \beta L_{\text{class}}(C_{\text{pred}}, C_{\text{gt}})$$

\subsection{\name's Dataset }
\label{sec:data}
Because of our unique radar point cloud processing pipeline, there are no available radar datasets that we can use to train and evaluate our model. We build a real-world data collection platform using two TI MMWCAS radars~\cite{ticascade} and a depth camera ZED 2i~\cite{zed2} as well as a phone equipped with multiple cameras and a LiDAR sensor, as elaborated in \ref{sec:imp}.   
However, due to the amount of data we need to train our network, depending on real-world data collection alone is impractical. We additionally generate our own augmented dataset that closely resembles the behavior and noise of the real-world radar data. This data consists of two parts; the first is simulated mmWave radar data, and the second is an augmented dataset inspired by the PCN~\cite{pcn} dataset, but is deformed to resemble radar noise patterns and practical motion perspectives. 

\vskip 0.05in
\noindent \textbf{(1) Radar Simulation.} 
We generate simulated data points by using a ray-tracing simulator provided in~\cite{guan2020hawkeye} and adapted for our radar and setup, namely two orthogonal 1-D virtual arrays. We capture 8 experiments per object in the dataset; in each experiment, the object is randomly positioned and oriented in the field of view. Moreover, we adjust the random orientations such that they are around the elevation axis, so that all of the experiments recover parts of objects that are typically observed from the perspective of a car, i.e., no flipped object. Finally, we use true-to-scale objects in the simulation set-up to ensure maximal compatibility with the real radar data. 
Each experiment consists of a number of frames, for each frame we simulate the horizontal and vertical heatmaps. Between frames, the radar antenna locations are changed in a linear trajectory parallel to the object to capture only one side of the object.

\vskip 0.05in
\noindent \textbf{(2) Geometric Perturbation Augmentation.}
One drawback of simulation is that due to the complexity of the simulator design, the simulation time for these experiments can take a long time. This led to the need for additional data that is faster to generate. The simulator accurately resembles the noise patterns and ambiguities associated with radar data, such as sinc noise. However, it does not properly capture random noise generated from real scenarios and hardware limitations. 

Thus, we use additional synthetic data generated from 3D models. We first randomly position a 'camera' around the 3D object, the camera should have a point of view similar to what a car would have as well. A snapshot is taken from this perspective, converted to a depth map, and sampled to generate the partial point cloud. This produces an ideal partial point cloud representation, which is repeated 8 times to generate 8 data points from different perspectives. We augment this representation to resemble radar data in two steps; the first step is to generate noise. As explained, mmWave radars have many artifacts in addition to their low angular resolution. We augment the partial point clouds to resemble these artifacts by replacing every point in the partial point cloud with multiple points randomly sampled from the surface of a sphere with a random, limited, radius that is centered around that point. This step creates noisier points that are slightly more dense, similar to our setup. The second step is to emulate specularity. As explained, specularity leads to the radar reflections missing some parts of objects, and while we partially overcome this challenge via temporal fusion, a practical trajectory will still suffer from specular signals since it cannot capture the object from every angle. To this end, we crop out multiple small spheres of the partial point cloud to emulate these parts as missing from the reflected signal. These spheres are randomly positioned so that at least one is near the center of the partial point cloud. This is based on the assumption that the center of most of our objects is the most specular part. 
The ground truth data are generated by uniformly sampling points from the 3D surface of the objects to represent the full shape.

\vskip 0.05in
\noindent \textbf{(3) Real-World Dataset.} 
We create a real-world radar dataset with 162 cars, 91 bikes, and 52 humans.
We plan to release the full dataset in the format of processed heatmaps and 3D point clouds, as well as the corresponding positional information per frame.
Our data collection experiments try to emulate a practical scenario as on self-driving cars by moving in mostly straight lines and observing objects only from one side, similar to what a typical car trajectory would be in reality. 
The trajectories of our partial point cloud experiments are short, straight segments with the radars facing in the general direction of an object of interest. The average trajectory length is about 3m, with a radar frame rate of approximately 60 fps. A single data point typically consists of 300\textasciitilde400 frames spanning about 6 seconds.

\section{Implementation and Setup}
\label{sec:imp}

\vskip 0.06in \noindent \textbf{Experimental Setup:} Our experimental setup shown in figure \ref{fig:hv_fusion} consists of two TI MMWCAS radars~\cite{ticascade}, which contain 12 transmitters and 16 receivers. Moreover, we use a depth camera ZED 2i~\cite{zed2} that is deployed on the same moving platform as the two radars to enable careful tracking of experiments. The depth camera also contains an inertial measurement unit and positional tracking that we use to localize the radars at each captured frame. The two radars and camera are synchronized at the software level to trigger the capturing of the three sensors at the exact moment via a handshake. We also use a mobile phone to scan the full ground truth shapes of all test objects via the PolyCam~\cite{polycam}, which utilizes the LiDAR and multiple cameras on the device. Our test experiments are done on static objects in different environments (indoor, outdoor) and different visibility conditions. 

 \vskip 0.05in
\noindent \textbf{Interference-Free Dual-Radar Operation.}
To prevent interference between the two radar sensors, we carefully design the FMCW frame configurations so that the radars do not transmit simultaneously. This is achieved by alternating the transmission of chirps between the two radars within a single frame. Specifically, in a frame containing $2N$ chirps, Radar 1 transmits and receives signals during the first $N$ chirps and remains silent for the remaining $N$ chirps. Conversely, Radar 2 remains silent during the first $N$ chirps and transmits during the second $N$ chirps. Experimental results confirm that any minor inconsistencies in chirp separation, caused by delays in the software handshake, are effectively mitigated by the low-pass filters integrated into the radar hardware. This ensures reliable operation without interference.

\vskip 0.05in
\noindent \textbf{Odometry.} 
There are several sensors that can be utilized for this purpose, including IMU, rotary encoder, or even camera-based Visual-Inertial odometry. Moreover, recent work employs the usage of mmWave radar heatmaps to extract accurate position information and perform SLAM~\cite{hong2020radarslam}. To reduce complexity, we extract the positional tracking information from the co-located depth camera placed on the platform for ground truth imaging.

\vskip 0.06in \noindent \textbf{Training \name:}
We use the augmented data to train our point cloud completion model. 
We use synthetic 3D CAD models from ShapeNet\cite{shapnet} objects dataset as well as CAESAR\cite{spring} human dataset to generate our dataset. 
Our dataset consists of 2,573 unique objects, including 1,697 cars, 396 bikes and motorbikes, and 480 human objects.
From each object, we generate 8 simulated and 8 synthetic data points using the simulator and synthesizing method explained in section~\ref{sec:data}.
Therefore, the total number of training data points is 41,168. 
For the ground truth, we sample the surface of the full 3D shapes uniformly to extract 16384 points, which the equal to the number of output points produced by the network.

We provide two versions of our model for comparison, \name\ and \name-no-bbox. Both models are trained on our generated dataset. \name\ is trained with the bounding box priors, such that the partial inputs are positioned in their correct position in the complete shape, and the object orientation is presumed to be fixed. On the other hand, \name-no-bbox is trained on the true-to-scale and centered data with diverse orientations in the training data. Both models have the same architecture and use EMD to measure coarse loss and CD for fine loss. We train our model for 340 epochs. We consider the following: 

\begin{itemize}
    \item \textit{Known Object Bounding Boxes.} We compare our results against three point-cloud reconstruction methods. 
For \name\ and each baseline method, we train an impractical model assuming the bounding-box is known in the pre-processing. We also train a practical model that does not require knowledge of input bounding boxes. 

\item \textit{Class-Specific Models.} We evaluate \name\ trained on all classes against three, class-specific \name\ models that are trained on only a single class each. This is to consider in the case of provided bounding boxes, it would be achievable to infer the classes of objects based on the bounding boxes. One can then use class-specific shape generation models to complete the shapes.

\item \textit{Fine-tuning}
We fine-tune \name\ with a portion of the real radar dataset to evaluate improved performance and reducing any simulation biases. We use 80\% of the collected data for fine-tuning and 20\% for testing. We do not include any objects seen in training for testing. 

\end{itemize}

\section{Experimental Evaluation}
\label{sec:results}

\begin{figure*}[t!]
    \centering
    \includegraphics[width=1.1\textwidth]{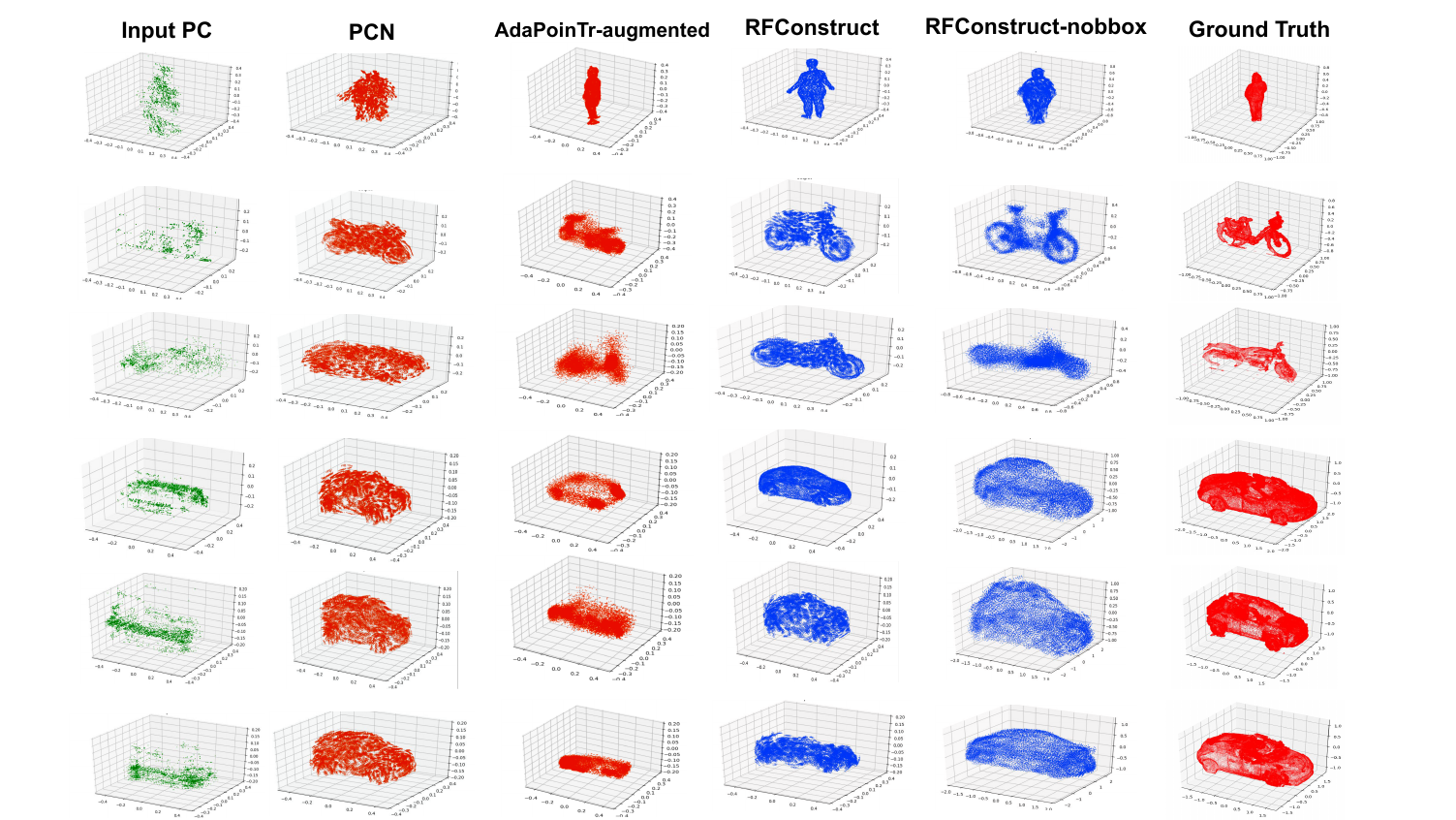}
    \caption{Performance against baselines using 3D radar point clouds without fine-tuning}
    \label{fig:baselines}
\end{figure*}

\subsection{Baselines}
 We compare our results against: 
 \begin{itemize}
     \item \textit{AdaPoinTr}~\cite{yu2023adapointr} is a point-cloud completion network with state-of-the-art performance; it is trained on 8 classes, including cars. We evaluate this model using their pre-trained weights and our testing data. 

     \item \textit{AdaPoinTr-augmented}~\cite{yu2023adapointr} We also train AdaPoinTr on our augmented dataset to evaluate the improvement of using our augmented dataset without any other changes on model architecture.

     \item \textit{PCN} ~\cite{pcn}. We also evaluate our results against the original PCN trained on the three classes of interest using the PCN data generation method. 
 \end{itemize}

\subsection{Metrics}
We use the common metrics to evaluate the difference between \name's output and the ground truth:
\begin{itemize}
    \item \textit{Chamfer Distance (CD)} is the minimum distance between a point in one point set and the closest point in the other set, the Chamfer distance quantifies the dissimilarity between two point clouds, facilitating a comprehensive assessment of the system's performance. 

    \item \textit{Earth Mover's Distance (EMD)}~\cite{fan2017emd} finds the one-to-one correspondence of points in the two point clouds and calculates the distance between each pair of associated points.

\end{itemize}

\subsection{Performance Against Baselines}

\subsubsection{Known Object Bounding Boxes}
We show \name's performance in comparison with the baselines in Fig.~\ref{fig:baselines} as well as in Table \ref{table:stacked}.
\name\ accurately reconstructs the 3D shapes of cars, bikes and humans that closely match the ground truth. In the reconstructed shapes, one can even clearly see the key defining features of different types of objects such as the wheels of cars and bikes, the arms and legs of humans, and even the seats and handles of bikes.
When having access to the 3D bounding boxes of objects and preprocessing the input point clouds as PCN~\cite{pcn}, \name's 3D reconstruction performance is comparably better than the PCN baseline by $1.06cm$ in CD and $1.3cm$ in EMD. Moreover, \name\ outperforms the AdaPoinTr baseline by  $0.6cm$ in CD and $5.8cm$ in EMD. 
Furthermore, note that although \name\ is trained primarily on simulated and synthesized data, it could generalize well to real scenes with different backgrounds and visibility conditions with only a small amount of fine-tuning as will be discussed below.

\subsubsection{Unknown Object Bounding Boxes}
As shown in Table \ref{table:stacked}, \name\ consistently outperforms baselines without bounding box priors, the baselines suffer from a large degradation in performance unlike \name\ which is capable of generating 3D reconstructions even without bounding box priors. \name-nobbox outperforms the PCN baseline by $14cm$ in CD and by $8.1cm$ in EMD. Moreover, \name-nobbox outperforms the AdaPoinTr baseline by  $14cm$ in CD and $8.7cm$ in EMD. 
As seen in Fig.~\ref{fig:baselines}, in the absence of 3D boundary boxes \name-nobbox can still successfully reconstruct the 3D shapes, whereas the PCN baseline simply fails.
It is important to note that when evaluating shapes generated without bounding box priors, the shapes are true to scale, where the average length of a car can be 2-3 meters. However, for shapes with bounding box priors, \textit{all} the complete shapes must fit within a 1 m sphere. This explains the difference in CD and EMD measurements between the two cases. 

\begin{table*}[t!]
\centering
\caption{Performance against baselines with and without object bounding box knowledge. CD and EMD reported are in cm.}
\resizebox{\textwidth}{!}{
\begin{tabular}{|l|c|c||l|c|c|}
    \hline
    \multicolumn{3}{|c||}{\textbf{With Bounding-box Priors}} & \multicolumn{3}{c|}{\textbf{Without Bounding-box Priors}} \\
    \hline
    & \textbf{Mean CD} & \textbf{Mean EMD} & & \textbf{Mean CD} & \textbf{Mean EMD} \\
    \hline
    AdaPointTr & 4.10 & 17.5 & AdaPointTr & 43 & 91.46 \\
    \hline
    PCN & 4.58 & 13.03 & PCN & 42.1 & 90.8 \\
    \hline
    \name & \textbf{3.52} & \textbf{11.7} & \name & \textbf{28.8} & \textbf{82.7} \\
    \hline
    Fine-tuned \name & \textit{\textbf{2.01}} & \textit{\textbf{6.25}} & Fine-tuned \name & \textbf{15.3} & \textbf{49.9} \\
    \hline
\end{tabular}
}
\vskip 0.1in
\label{table:stacked}
\end{table*}

\subsubsection{Real Data Fine-tuning} \label{sec:finetune}
We fine-tune \name\ on a portion of the real radar dataset we collected in an effort to bridge the gap between the synthetic training data and real-world objects and noise. We show that fine-tuning improves the performance of \name\ by $2.5 cm$ in CD and $5.45 cm$ in EMD when using bounding box priors, and by $13.5 cm$ in CD without any knowledge about bounding boxes, as shown in table \ref{table:stacked}. We also compare visually the results from fine-tuning with the basic \name\ in figures \ref{fig:random_1}, \ref{fig:random_2}, and \ref{fig:random_nobbox}.

\subsubsection{Class-Specific Models.}
We evaluate \name\ trained on all classes against three, class-specific \name\ models that are trained on only a single class each. This is to consider in the case of provided bounding boxes, it would be achievable to infer the classes of objects based on the bounding boxes. One can then use class-specific shape generation models to complete the shapes. However, as shown in Table \ref{table:class_performance}, the improvement of a class-specific model over our general model is marginal and inconsistent. This is likely due to the inherent classification abilities of \name. In addition, the model generalized well by training on more data for all classes.

\subsection{Ablation Studies}
\label{sec:ablation}

\subsubsection{Ablation of the Classifier Module}
\label{sec:classifier_ablation}

The classifier module aids the encoder in generating features that are class-separable. This in turn improves the decoder performance due to having more representative input features. This simply reduces the instances of generating a compete shape of the wrong class based on the partial point cloud input. The addition of the classifier module improves the overall CD and EMD as can be seen in table  \ref{table:ablation_combined}. We analyze the impact of the classifier module by removing it from the pipeline and observing the changes in performance.

\subsubsection{Ablation of Data Augmentations}
\label{sec:augmentation_ablation}

\name\ is trained on a merged dataset of simulated radar data and synthetic data. This enables the network to learn radar data features, in addition to adapting to noise and variance present in real measurements. Training only on one of these data types results in worse performance as shown in table \ref{table:ablation_combined}. Training or a combination of different data improves the CD and EMD.  The role of data augmentations in improving performance is studied by training the model without them.
To further test the effectiveness of our training dataset, we trained AdaPoinTr on our \name\ dataset. This resulted in improved performance by $2.6cm$ in mean EMD. The improvement from leveraging both kinds of augmentations is likely due to their combined resemblance to the radar point clouds. While the simulated data models the noise of radar systems such as low resolution, $sinc$ noises, and grating lobes, the geometric perturbations can model signal specularity as well as noise generated by real experiments and systems.

\vskip 0.3in
\begin{minipage}{0.42\textwidth} 
    \centering
    \resizebox{\columnwidth}{!}{
    \begin{tabular}{|l|c|c|}
    \hline
    \multicolumn{3}{|c|}{\textbf{Classifier module}} \\ \hline
    Configuration & CD & EMD \\ \hline
    With Classifier Module & \textbf{3.52} & \textbf{11.7} \\ \hline
    Without Classifier Module & 4.2 & 12.3 \\ \hline
    \multicolumn{3}{|c|}{\textbf{Data augmentations.}} \\ \hline
    Training Data & CD  & EMD  \\ \hline
    Combined Augmentations & \textbf{3.52} & \textbf{11.7} \\ \hline
    Simulated Radar Data Only & 4.62 & 14.1 \\ \hline
    Geometric Augmentation Only & 4.09 & 12.53 \\ \hline
    AdaPointTr & 4.10 & 17.5 \\  \hline
    Augmented AdaPoinTr & 3.42 & 14.94 \\  \hline
    \end{tabular}
    }
    \captionof{table}{Comparison of results with and without the classifier module and data augmentations.}
    \label{table:ablation_combined}
\end{minipage}%
\hfill 
\begin{minipage}{0.48\textwidth}
    \centering
    \resizebox{\columnwidth}{!}{
    \begin{tabular}{|c|c|c|c|}
        \hline
        Classes in & Classes in & CD $\downarrow$ & EMD $\downarrow$ \\
        Evaluation & Training & (cm) &  (cm)  \\
        \hline
        \multirow{2}{*}{Human} & Human only  & 2.8 & 8.34  \\
        \cline{2-4}
        & Car, Bike, Human & 3.1 & 11.5  \\
        \hline
        \multirow{2}{*}{Car} & Car only & 2.65 & 8.66 \\
        \cline{2-4}
        & Car, Bike, Human & 3.1 & 10.3  \\
        \hline
        \multirow{2}{*}{Bike} & Bike only & 4.3 & 13.3 \\
        \cline{2-4}
        & Car, Bike, Human & 4.2 & 13.6  \\
        \hline
    \end{tabular}
    }
    \captionof{table}{Class-specific models performance}
    \label{table:class_performance}
\end{minipage}

\begin{figure}[!t]
    \begin{minipage}[t]{0.5\textwidth} 
        \centering
        \includegraphics[trim=0in 0in 0in 0.3in, width=\linewidth]{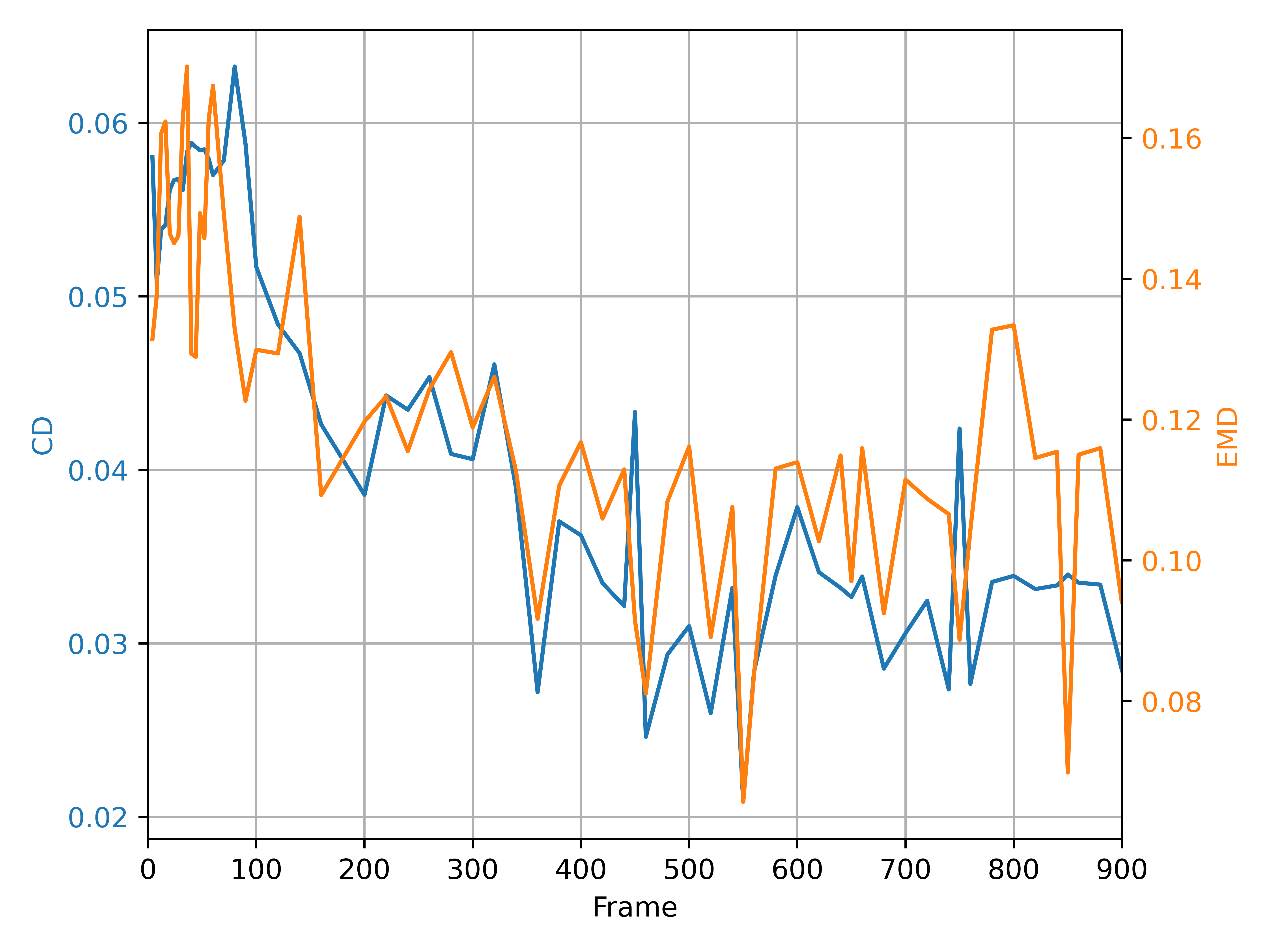}
        \caption{Reconstruction performance against input point cloud with increasing number of merged frames.}
        \label{fig:frame_ablation}
    \end{minipage}
    \hfill 
    \begin{minipage}[t]{0.48\textwidth} 
          \centering
        \includegraphics[trim=1.8in 0.0in 3.9in 0in, clip, width=\linewidth]{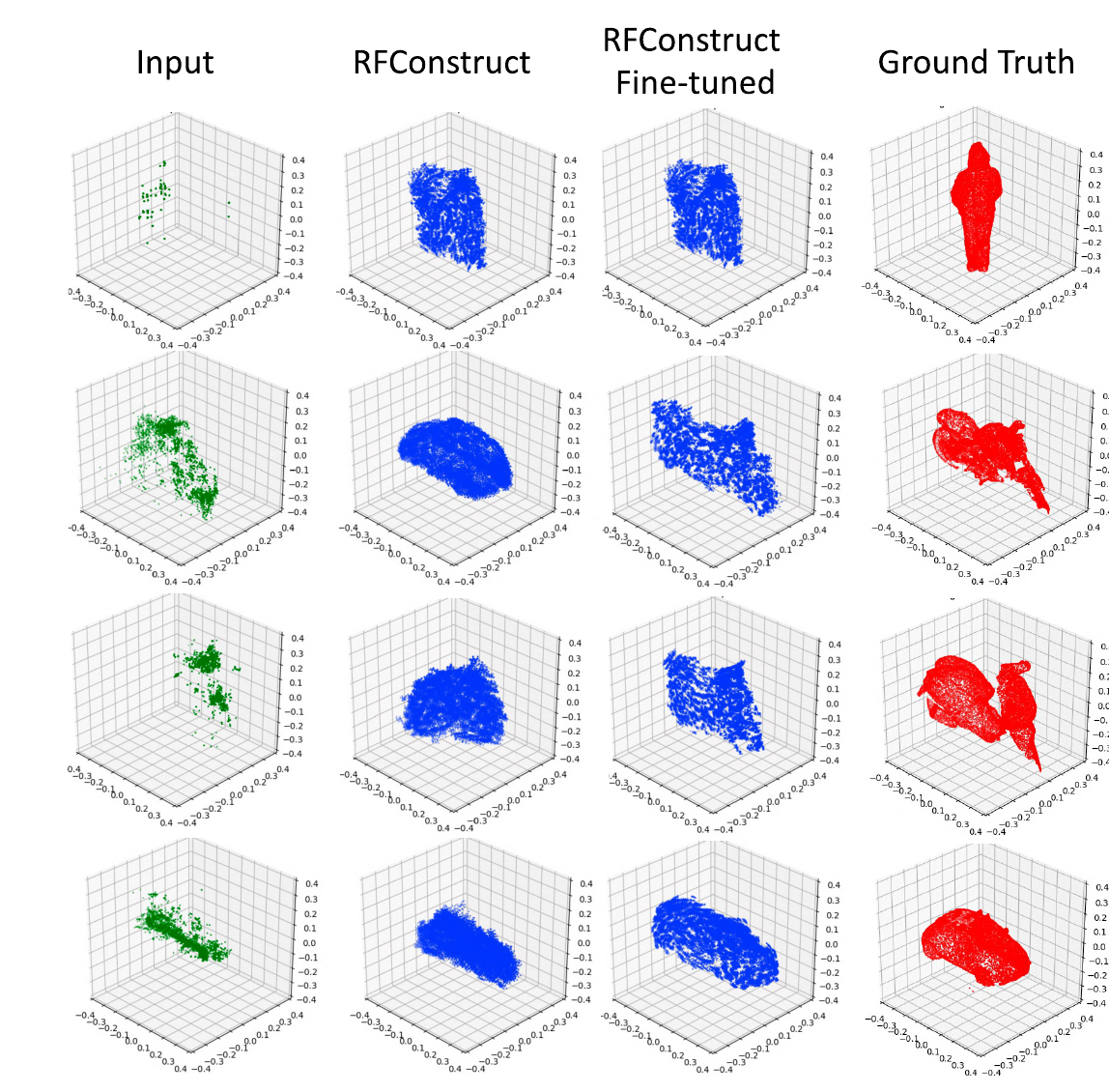}
    
        \caption{\name\ failure cases}
        \label{fig:failure}
    \end{minipage}
    \end{figure}
\subsection{Impact of Trajectory on Reconstruction Quality}

\label{sec:trajectory_length}

We study the effect of increasing the trajectory length (number of frames) on the quality of the reconstructed shapes. Results indicate that longer trajectories generally improve completion quality. However, the value gained by adding more frames decreases as the total number of frames increase, as can be shown in figure \ref{fig:frame_ablation}. 

The plateau in the reconstruction improvement can be due to the practical, linear, trajectories in which most of our experiments are conducted, where additional frames contribute fewer information about the object. As the trajectory progresses, new frames capture increasingly redundant information about the object's geometry. This redundancy arises because linear trajectories are limited to a specific viewpoint. Consequently, the marginal improvement in shape completion decreases with each additional frame.

\subsection{SAR vs \name\ combination}

As discussed in Section \ref{sec:Background}, Synthetic Aperture Radar (SAR) can combine radar information captured from different locations by utilizing scanning antennas. This approach improves resolution by emulating large antenna arrays using smaller ones. However, SAR is highly sensitive to small errors in antenna locations since half a wavelength at 77 GHz is less than 2 mm. Hence, we need millimeter level accuracy, otherwise the error in the phase of the signal leads to the accumulation of sinc noise rather than its cancellation.

\begin{figure*}[t]
    \centering
    \includegraphics[width=\textwidth]{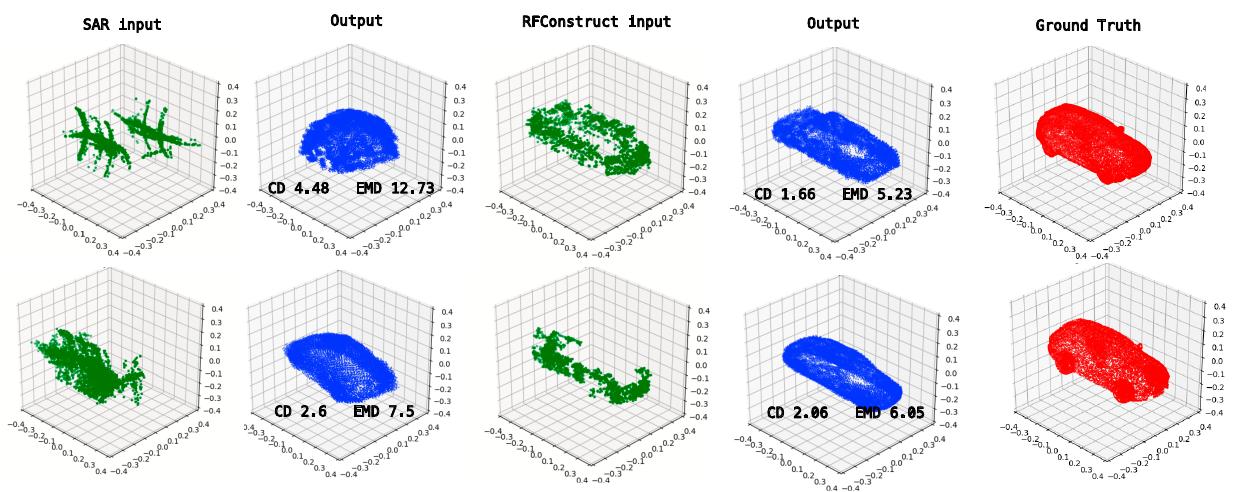}
    % \vskip -6in 
    \caption{Comparing SAR point-clouds to \name\ point-clouds and completion performance. }
    \label{fig:sar}
    % \vskip 3.2in
\end{figure*}

We conduct experiments with SAR imaging using our experimental setup. As shown in Fig\ref{fig:sar}, the coherent combination of the collected frames amplifies sinc noise and results in prominent grating lobes. In contrast, our proposed combination method produces accurate, though sparse, point clouds. Furthermore, when both point clouds—from the coherent combination and our \name\ method—are used as inputs to complete the 3D shape, the output derived from the \name\ point cloud is both quantitatively and qualitatively better than that obtained from the coherent combination. In particular, the CD can decrease from 4.48 to 1.66 and the EMD from 12.73 to 5.23.

\subsection{Failure Cases}
 
We were able to identify some specific conditions under which our system fails to provide correct reconstructions, as can be seen in figure \ref{fig:failure}. One case is when the input point cloud is exceedingly sparse, providing insufficient geometric information for accurate reconstruction. With too few points, the model struggles to infer meaningful structure, leading to incomplete or incorrect shapes. Failure can also occur as the Signal-to-Noise ratio of the input point cloud is very low, where noise points significantly outnumber valid ones. In this case, the model could mistakenly identify some of the noise points as valid points that represent the geometry of the object. This results in incorrect or distorted shapes. Addressing these limitations could involve incorporating mechanisms to handle sparse data more effectively and improve robustness to noisy inputs, such as advanced denoising techniques or prior knowledge of object structure.

Finally, the gap between the simulated training data and real data could lead to some misrepresentation of the object's geometry during reconstruction. This gap arises because the simulated training data often lacks the complexities and imperfections present in real-world scenarios, and are difficult to manufacture, such as measurement noise, environmental interference, or material properties that affect reflected signals. Which is why fine-tuned \name\ can provide a better reconstruction in some of the failure cases shown in figure \ref{fig:failure}.

\subsection{Additional Results from Randomly Selected Data Points}
\label{sec:more_res}

\begin{figure*}[t!]
    \centering
    \includegraphics[page=1, trim=1.5in 0in 2in 0.0in, clip, width=\textwidth]{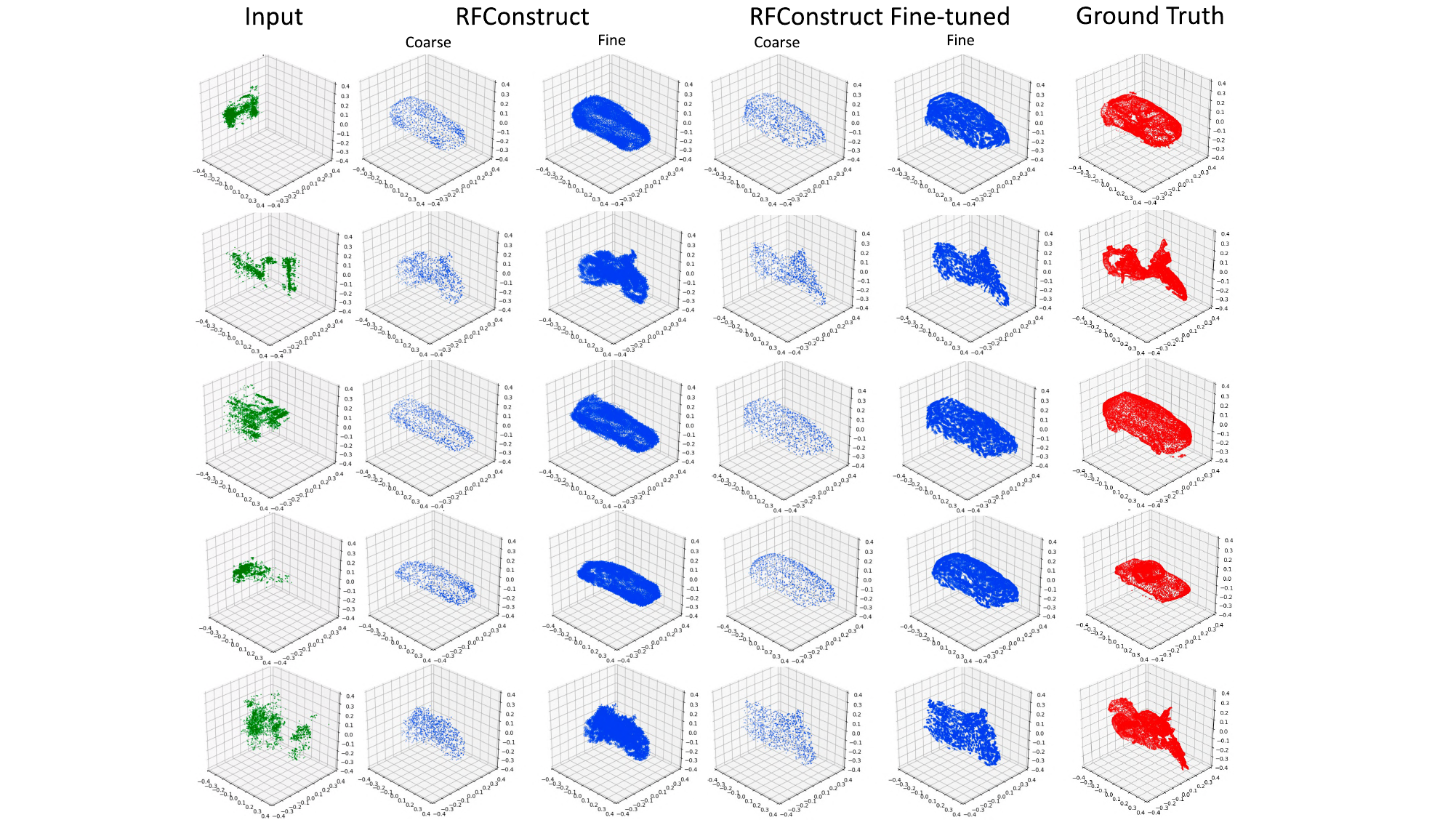}
    \vskip 0.1in 
    \caption{Qualitative results for \name\ and \name\ fine-tuned on real radar data for randomly selected data points. }
    \label{fig:random_1}
    \vskip -0.2in
\end{figure*}

\begin{figure*}[b]
    \centering
    \includegraphics[page=1, trim=1.5in 6.99in 2in 0.0in, clip, width=\textwidth]{results_color_1.pdf}
    \includegraphics[page=1, trim=1.5in 0in 2in 0.1in, clip, width=\textwidth]{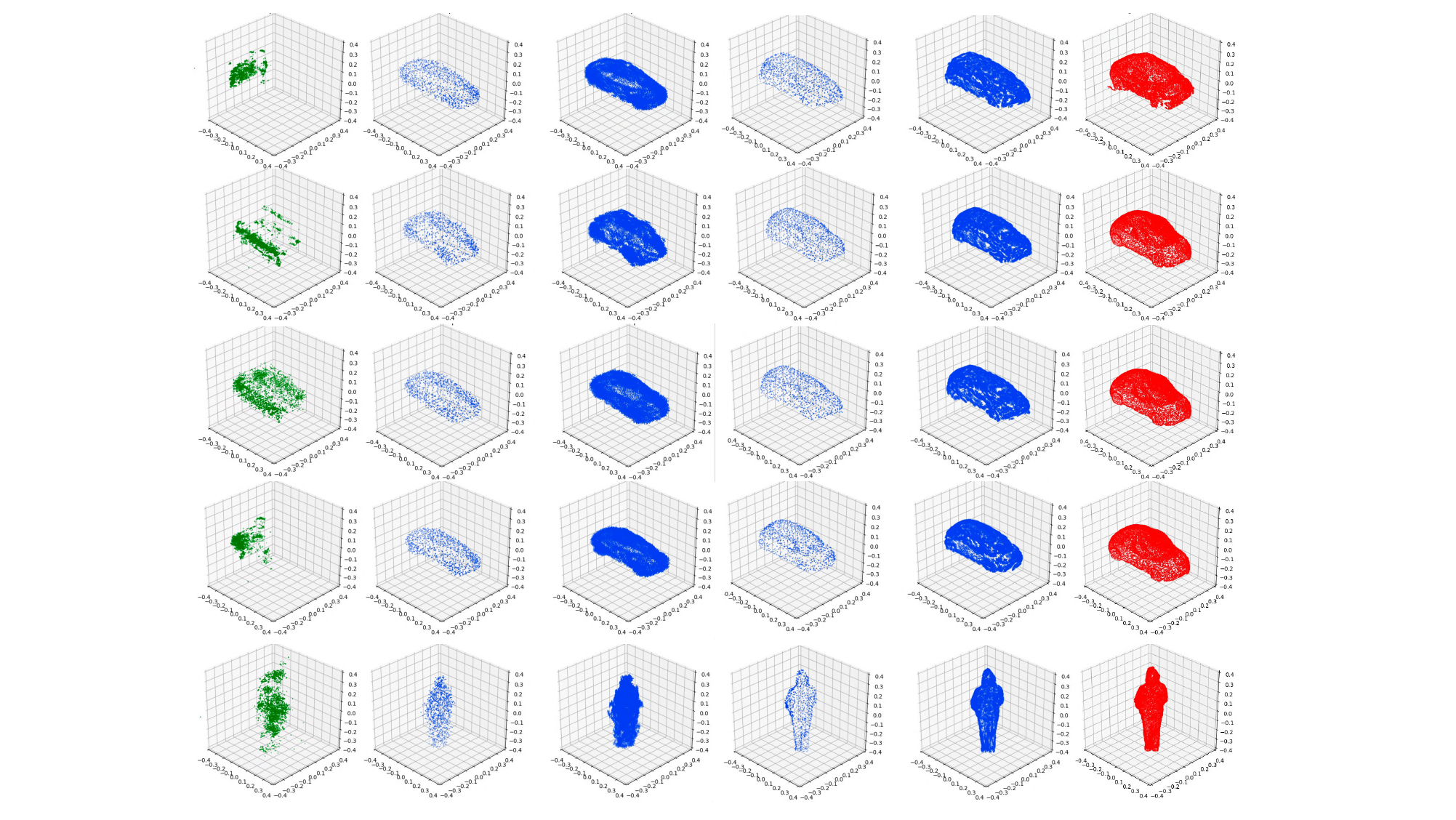}
    \includegraphics[page=1,trim=1.5in 6in 2in 0.15in, clip , width=\textwidth]{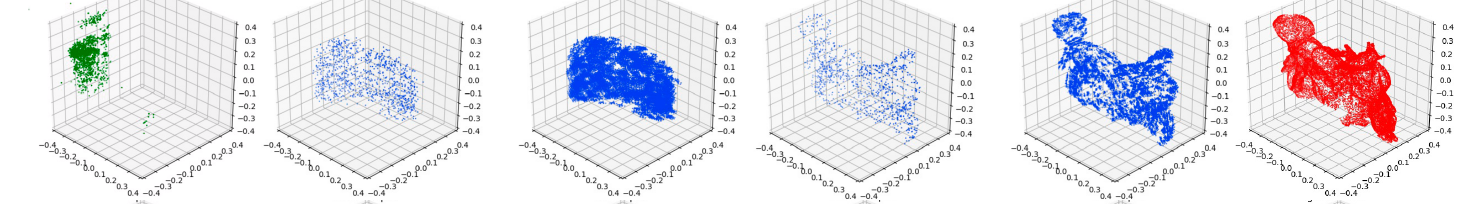}
    \vskip 0.1in 
    \caption{Qualitative results for \name\ and \name\ fine-tuned on real radar data for randomly selected data points.}
    \label{fig:random_2}
    \vskip -0.3in
\end{figure*}

\subsubsection{\name\ with Bounding-Box Priors}
In figures \ref{fig:random_1} and \ref{fig:random_2} we show results for \name\ and fine-tuned \name\ for randomly selected data points from the testing set. We show both the coarse and fine outputs from the network. While the basic system without fine-tuning produces representative reconstructions for most cases, Fine-tuning improves the reconstructions and brings the representations much closer to the actual ground truth shapes.

\subsubsection{\name\ without Bounding-Box Priors}
\begin{figure*}[b!]
    \centering
    \includegraphics[page=1, trim=2in 0.5in 1in 0.5in, clip, width=1\textwidth]{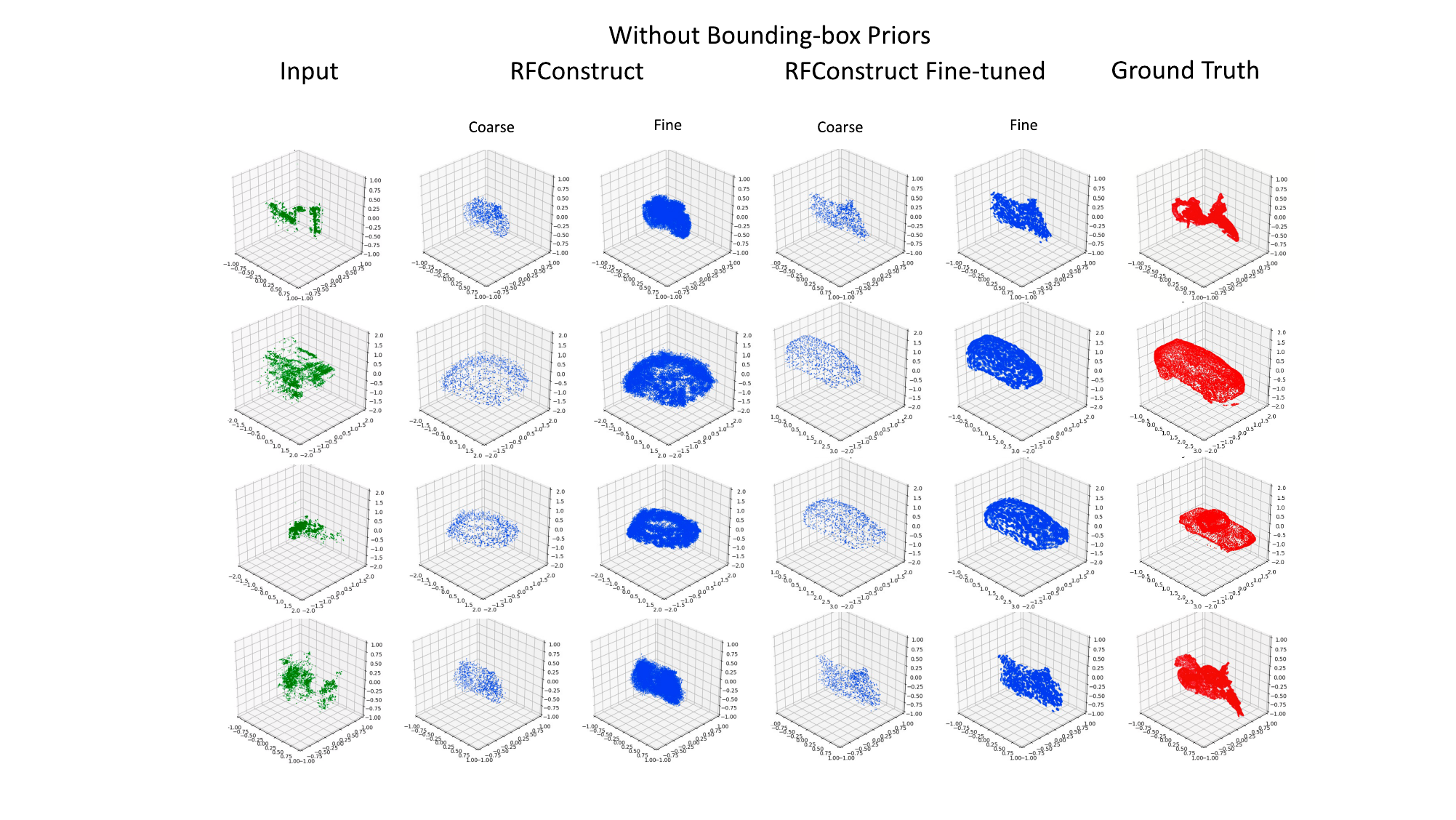}
    \includegraphics[page=1, trim=2in 0in 1in 2.96in, clip, width=1\textwidth]{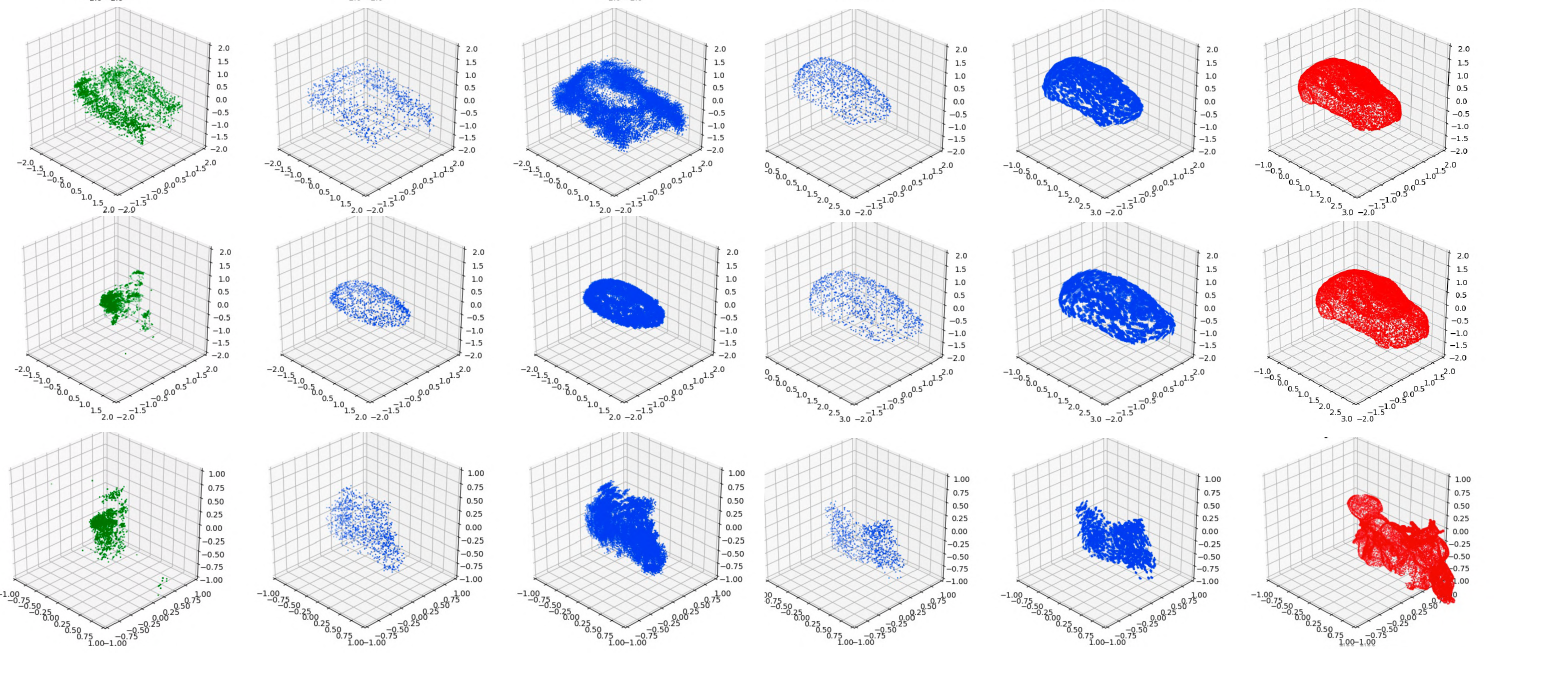}
    \caption{Qualitative results for \name\ and \name\ fine-tuned on real radar data \textbf{without} bounding-box priors for randomly selected data points.}
    \label{fig:random_nobbox}
    \vskip -1in
\end{figure*}

In the case of testing without bounding box priors, the reconstruction is more challenging as more ambiguity arises about the object type, orientation and dimensions. In figure \ref{fig:random_nobbox}, we show random samples of the data tested with \name\ and fine-tuned \name\ trained without bounding box priors. 

\section{Limitations \& Future Work}
\label{sec:limitations}

While \name takes major steps towards enabling 3D shape reconstruction from mmWave radars in autonomous driving, it still has several limitations that require addressing before it can be fully realized in practice.

\begin{itemize}
    \item {\it Moving targets:}
Our current implementation of \name only works for static objects. Imaging moving objects requires significantly more research as we cannot easily fuse temporal radar frames. We will need to develop algorithms to estimate and compensate for the relative motion and relative positions of the objects with respect to the ego car. We will also explore the tradeoff between smearing the object and combating specularity in setting the temporal fusion window, which depends on the relative speed and relative direction of motion. 

    \item {\it Generalizing to other shapes \& environments:} We only trained and tested \name for three classes of objects that represent the most common moving objects on the road that present safety risks. To make \name more general, we need to extend it to new classes like trees, road signs, fire hydrants, trash cans, large trucks, etc. We also need to update the neural network architecture and training to generalize to different streets and roads with different building structures. 

\item {\it Incorporating Doppler: } \name currently does not leverage the Doppler information from radar. Incorporating Doppler would allow us to separate nearby objects along the Doppler domain (cars, bikes, and pedestrians move at different speeds). Doppler also allows us to estimate the speed of various objects in the scene, which is essential for fusing temporal radar frames of moving objects.

    \item {\it Pedestrian posture:} While \name can recover the orientation of the cars and bikes in the scene, it cannot estimate the orientation or posture of pedestrians. Currently, it only estimates a rough silhouette but cannot accurately recover the human posture, which can be useful for driving cars to estimate the direction the pedestrian is walking. We can build on the extensive literature on human pose estimation~\cite{zhao2018rfpose, zhao2018rfpose3d, xue2021mmmesh, chen2022mmbody,xie2023mmPoint, kong2022m3track, lee2023hupr} from radar signals and incorporate it into \name.  
\end{itemize}

\section{Conclusion}
\label{sec:conclusion}
This paper takes major steps towards enhancing autonomous radars by developing a radar system that provides good resolution in both azimuth and elevation and completes 3D shapes of important objects in the context of autonomous driving, such as cars, bikes, and pedestrians, from partial observations. Such capabilities will be essential in enabling self-driving cars to operate in fog and bad weather. \name\ was able to tackle challenges in hardware, modality, and practicality to enhance performance. However, significant more research is needed to enhance the robustness, stability, and generalizability of such systems.

{
    % \small
    \bibliographystyle{ACM-Reference-Format}
    \bibliography{main}
}

\end{document}